\documentclass[aps,prb,floats,floatfix,amssymb,amsmath,prb,twocolumn,superscriptaddress]{revtex4-2}
\usepackage[english]{babel}
\usepackage{amsmath}
\usepackage{bm}
\usepackage{graphicx,bbm} 
\usepackage{times}
\usepackage{epsfig} 
\usepackage[normalem]{ulem}
\usepackage[utf8]{inputenc}
\usepackage[colorlinks,linkcolor=blue,citecolor=blue,urlcolor=blue]{hyperref}
\usepackage{color}
\usepackage{latexsym}
\begin{document} 
\title{Spin and energy diffusion vs. subdiffusion in disordered spin chains}
\author{J. Herbrych}
\affiliation{Institute of Theoretical Physics, Faculty of Fundamental Problems of Technology, Wroc\l{a}w University of Science and Technology, 50-370 Wroc\l{a}w, Poland}
\author{P. Prelov\v{s}ek}
\affiliation{Jo\v{z}ef Stefan Institute, SI-1000 Ljubljana, Slovenia}

\date{\today}
\begin{abstract}
While the high-temperature spin diffusion in spin chains with random local fields has been the subject of numerous studies concerning the phenomenon of many-body localization (MBL), the energy diffusion in the same models has been much less explored. We demonstrate that energy diffusion is faster at weak random fields but becomes essentially equal at strong fields; hence, both diffusions determine the slowest relaxation time scale (Thouless time) in the system. Numerically reachable finite-size systems reveal the anomalously large distribution of diffusion constants with respect to actual field configurations. Despite the exponential-like dependence of diffusion on field strength, the results for sensitivity to twisted boundary conditions are incompatible with the Thouless criterion for localization and the presumed transition to MBL, at least for numerically reachable sizes. In contrast, we find indications of the scenario of subdiffusive transport, particularly in the dynamical diffusivity response.
\end{abstract}

\maketitle

\section{Introduction}
The localization and absence of diffusion in systems of noninteracting fermions are by now well-established phenomena since the original proposal \cite{anderson58}, with well-understood scaling properties of the related diffusion-localization transition, both explained theoretically \cite{abrahams79,evers08} as well as confirmed numerically (for a review see \cite{kramer93,markos06}). In contrast, the introduction of interaction and the corresponding problem of many-body localization (MBL) \cite{basko06} remains a hard challenge despite nearly two decades of intensive studies. Still, several qualitative features and markers of the MBL regime have been well identified and established, predominantly via numerical studies of the 'standard' model of MBL, i.e., the anisotropic Heisenberg chain with random local fields. At large field strength $W>W^*$ typically one finds the change in level and spectral statistics \cite{oganesyan07,luitz15,serbyn16,suntajs20,sierant20}, logarithmic growth of entanglement entropy \cite{znidaric08,bardarson12,serbyn15}, strong suppression of spin diffusion \cite{berkelbach10,barisic10,agarwal15,barlev15,steinigeweg16,prelovsek17}, and generally nonergodic correlations \cite{pal10,serbyn13,huse14,luitz16,mierzejewski16}. However, the fundamental question of whether the MBL remains stable in the thermodynamic limit and long times \cite{suntajs20,suntajs_bonca_20,sels2020,vidmar21,krajewski22}, and related question whether the MBL regime emerges either as the (dynamical) phase transition or more like glassy crossover at marginal $W\sim W^*(L)$ which can depend on the system size $L$, is still unanswered. 

So far, numerical studies of high-$T$ spin d.c. conductivity $\sigma^0_s$ and the corresponding spin diffusion constant ${\cal D}^0_s$ established exponential-like dependence on the disorder strength $W$ \cite{barisic10,barisic16,steinigeweg16,prelovsek17,prelovsek21,herbrych22}, which appears consistent with the related exponential increase of the Thouless time $\tau_{\mathrm{Th}}(W)$ \cite{suntajs20,suntajs_bonca_20}. However, due to the modest system length $L$ reachable numerically, results are heavily influenced by large random-field sample-to-sample variations \cite{herbrych22,vidmar21,krajewski22-2}. This contrasts with analogous transport results for Heisenberg chains with quasiperiodic fields \cite{prelovsek23}, where fluctuations are minimal. In such systems, one can also identify MBL-like behavior (e.g., exponential-like dependence of transport properties on disordered strength $W$) \cite{iyer13,barlev17,khemani17,setiawan17,znidaric18,zhang18,aramthottil21,sierant22}, being also realized experimentally in cold-atom systems \cite{schreiber15,luschen17}. Recently, the debate of the MBL regime has been revived by the analytical study indicating that even at large disorder, the transport is not localized but rather subdiffusive \cite{deroeck24}, having the origin in insulating/blocking regions and related Griffiths effects already considered in MBL models \cite{agarwal15,gopal16,luschen17}. 

In contrast to the spin diffusion ${\cal D}^0_s$ (or particle diffusion in the equivalent model of interacting spinless fermions), the energy diffusion ${\cal D}^0_e$ (and related thermal conductivity \cite{prelovsek17,jencic15}) in disordered spin chains remains relatively unexplored. While for a weak disorder, it is expected and also found to be much larger/faster \cite{mendoza19}, few studies of larger disorders yield conflicting messages \cite{varma17,schulz18}. The relation between both diffusions is relevant since the relaxation time to the equilibrium (referred as the Thouless time, also closely related to the concept of Thouless energy \cite{serbyn17}) is supposed to be given as by the longest $\tau_{\mathrm{Th}} \propto L^2/{\cal D}^0$ \cite{edwards72,suntajs20,sierant20,sierant24} being determined either by ${\cal D}^0_s$ or ${\cal D}^0_e$. To treat both diffusions on the same level, we derive the corresponding Einstein relation for ${\cal D}^0_e$, which allows us to follow numerically in finite systems the energy diffusion up to the same accuracy as ${\cal D}^0_s$. In the following, we show that at larger $W$, both diffusion constants are becoming nearly equal; hence both are relevant to $\tau_{\mathrm{Th}}$.

Due to the strong dependence of d.c. transport on particular random-field configurations, it is important to evaluate numerically the diffusion in reachable finite-size separately for individual samples and establish whether their properties (also as the test of the ergodicity) comply with the expected eigenstate-thermalization hypothesis (ETH) and random-matrix theory (RMT) universality \cite{wilkinson90,dalessio16}. We study this by the Thouless approach \cite{edwards72} relating the d.c. transport (or its absence, i.e., localization) to the flux (twisted-boundary) sensitivity. Instead of averaging over random filed configurations, we investigate distributions of diffusion values, which become very broad and far from standard Gaussian distribution, i.e., at larger $W$ distributions become closer to log-normal distribution \cite{herbrych22} with long tails which might be signatures of Griffiths effects and of the related subdiffusive scenario \cite{deroeck24}. Nevertheless, much clearer support for the subdiffusion emerges from the dynamical diffusion response which reveals ${\cal D}(\omega)\sim{\cal D}^0+c|\omega|^\alpha$ with $\alpha<1$ \cite{agarwal15,gopal16} in a wide range of $\omega$ at large $W$, which is contrast to case of quasiperiodic potential \cite{prelovsek23}, where we find systematically $\alpha \sim 1$. 

\section{Spin and energy diffusion}

We study the 'standard' model for MBL, i.e., a spin chain described by the XXZ Heisenberg model with random local fields,
\begin{equation}
H = \sum_l \left[ \frac{J}{2}( \mathrm{e}^{-i \varphi} S^+_{l+1} S^-_l+ \mathrm{H.c.}) 
+ J \Delta S^z_{l+1} S^z_l + w_l S_l^z \right]\,,
\label{hm}
\end{equation}
with $S=1/2$ spin operators and random fields \mbox{$w_l\in[-W,W]$} with the uniform probability distribution. We consider the chain with $L$ sites and periodic boundary conditions numerically. To allow the Thouless approach to test transport properties via sensitivity to (twisted) boundary conditions, as well as the validity of RMT universality, \cite{edwards72,prelovsek23}, we introduce in general a finite flux $\varphi\ne0$. As with most MBL studies, we fix for convenience in numerical studies the (Ising) interaction corresponding to the isotropic case $\Delta=1$ and use $J=1$ as the unit of energy. Within the MBL problem, we are interested in the dynamical transport properties at finite temperature $T>0$, and as usual, for simplicity, we deal with the $T\to\infty$ limit. While d.c. transport quantities (i.e., the spin conductivity $\sigma^0_{s}$ and thermal conductivity $\kappa^0$) vanish at $T\to\infty$, the related diffusion constants do not and remain highly nontrivial quantities to determine. 

The dynamical spin diffusivity at $T\to\infty$ and the Einstein relation to the (dynamical) spin conductivity $\sigma_s(\omega)$ can be conveniently derived \cite{mori65,forster75} via the dynamical susceptibility $\chi_q(\omega)$ and the related relaxation function $\phi_q(\omega)$, 
\begin{eqnarray}
\chi_q(\omega) &=& \frac{1}{\beta} \int_0^\infty \mathrm{d}t \mathrm{e}^{i \omega t}\,\langle [S^z_{-q}(t), S^z_{q}] \rangle\,, ~~~
S^z_q = \frac{1}{\sqrt{L}} \sum_l \mathrm{e}^{i q l} S^z_l, \nonumber \\
&&\phi_q(\omega) = \frac{\chi_q(\omega) - \chi^0_q}{\omega} = \frac{ - \chi^0_q}{ \omega +M_q(\omega) }\,.
\label{chiq}
\end{eqnarray}
In the hydrodynamic regime $q\to0$ \cite{forster75} $M_q(\omega) \sim iq^2\sigma_s(\omega)/\chi^0_q=iq^2{\cal D}_s(\omega)$, leading to the dynamical spin diffusion 
\begin{eqnarray}
{\cal D}_s(\omega) &=& \frac{\sigma_s(\omega)}{\chi_s^0} = \frac{\pi}{L \tilde{\chi}_s^0} \int_0^\infty \mathrm{d}t\,\langle j_s(t) j_s \rangle 
\nonumber \\
&=& \frac{\pi}{L \tilde{\chi}_s^0 N_{st}} \sum_{m\ne n} | \langle m|j_s |n \rangle |^2 \delta(\omega - e_m + e_n)\,.
\label{dsw}
\end{eqnarray}
In the last equation, ${\cal D}_s(\omega)$ is expressed directly in terms of spin-current matrix elements among many-body (MB) eigenstates $|n\rangle$, $|m\rangle$, and $N_{st}$ representing the total number of MB states. The spin current $j_s$ and high-$T$ static susceptibility $\chi_s^0=\chi^0_{q\to0}=\tilde{\chi}_s^0/T$ are within the considered model \eqref{hm} given by
\begin{equation}
j_s = -\frac{J}{2} \sum_l (i \mathrm{e}^{-i \varphi} S^+_{l+1} S^-_l+ \mathrm{H.c.})\,, \quad
\tilde \chi_s^0 = \langle S^z_{-q} S^z_q \rangle = \frac{1}{4}\,.
\end{equation}

While the above Einstein relation for spin diffusion, Eq.~\eqref{dsw}, has been derived and used before \cite{bonca95,prelovsek23}, it is convenient to follow the same steps for the energy density modulation $h_q$, to derive the expression for energy diffusivity ${\cal D}_e(\omega)$ 
\begin{equation}
h_{q} = \frac{1}{\sqrt{L}} \sum_l \mathrm{e}^{i q l} h_l, \qquad H= \sum_l h_l\,,
\end{equation}
arriving in the hydrodynamic regime $ q\to0$ to
\begin{eqnarray}
{\cal D}_e(\omega) &=& \frac{\sigma_e (\omega)}{\chi_e^0} = \frac{\pi}{L \tilde{\chi}_e^0} \int_0^\infty \mathrm{d}t\,
\mathrm{e}^{i \omega t} \langle j_e(t) j_e \rangle \nonumber \\
 &=& \frac{\pi}{L \tilde{\chi}_\epsilon^0 N_{st}} \sum_{m\ne n} | \langle m|j_e |n \rangle |^2 \delta(\omega - e_m + e_n)\,,
\label{dew}
\end{eqnarray}
with the energy current \cite{karahalios09} $j_e=\sum_l j_{e l}=i\sum_l [h_l,h_{l+1}] $ and static $T\to\infty$ energy susceptibility $\tilde{\chi}_\epsilon^0$ given by
\begin{eqnarray}
&&\tilde{\chi}_e^0 = \langle h_{-q} h_q \rangle = \frac{J^2}{8}+ \frac{J^2 \Delta^2}{16}+ \frac{W^2}{12}\,, 
\label{chie} \\
j_{e l} &=& - J S^z_l j_{l-1}^{l+1} + \left[J \Delta \left(S^z_{l+2}+ S^z_{l-1}\right) + \frac{w_l+w_{l+1}}{2}\right] ~ j_l^{l+1}\,, \nonumber \\
&&j_l^m = - \frac{J}{2} \left( i \mathrm{e}^{-i \varphi (m-l)} S^+_{m} S^-_l +\mathrm{H.c} \right)\,. \label {je}
\end{eqnarray}
Note that $j^m_l$ is related to the spin current $j_s=\sum_\ell j^{l+1}_l$.

\section{Testing results vs. random-matrix theory} 

Small systems up to $L\leq18$ sites with the dimension of the Hilbert space $N_{st}\sim5.10^4$ allow the application of the full exact-diagonalization (ED) of the Hamiltonian \eqref{hm}, and consequently, the explicit evaluation of matrix elements $j^{s,e}_{mn}=\langle m|j_{s,e}|n \rangle $, needed in Eq.~\eqref{dsw} and Eq.~\eqref{dew}. Moreover, at finite flux $\varphi\ne0$ (note that at $\varphi=0$ all diagonal elements vanish, i.e. $j_{nn}=0$), one can test whether the particular samples, i.e., finite-size systems with chosen field configurations $w_l$, satisfy the ETH universality, in analogy to the previous study of quasiperiodic potentials \cite{prelovsek23}. For the present case of random $w_l$, the universality in the ergodic regime and $\varphi\ne0$ is expected to obey RMT and the Gaussian orthogonal ensemble (GOE). In particular, one expects $Y=\overline{|j_{mn}|^2}/\overline{|j_{nn}|^2}\sim 1$ when the average is performed in the narrow enough range of eigenstates $|e_m-e_n|\to 0$. At the same time, fluctuations should be Gaussian, e.g., with the kurtosis $Q=\overline{|j_{nn}|^4}/(\overline{|j_{nn}|^2})^2=3$. Since the variations between different random samples are large even for intermediate $W$ \cite{herbrych22}, it is important to test RMT for individual $w_l$ configurations. 
 
\begin{figure}[!htb]
\includegraphics[width=1.0\columnwidth]{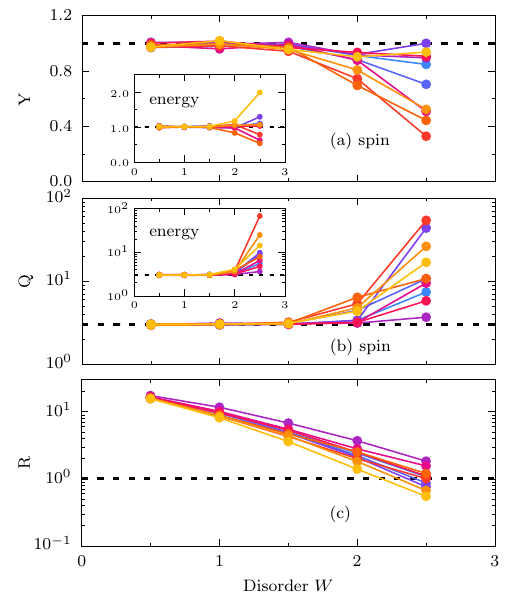}
\caption{RMT markers vs. random-field strength $W$ for the spin current matrix elements (a) $Y_s$, (b) $Q_s$, and corresponding energy current elements $Y_e$, $Q_e$ (as the insets) for $N_c=10$ different random-field $w_l=W\eta_l$ configurations on a system of $L=18$ sites. Lines $Y=1$ and $Q=3$ denote the expected GOE values. (c) The corresponding results for level sensitivity $R$ vs. $W$ with marked marginal value $R=1$.}
\label{fig1}
\end{figure}

For the spin transport the relevant criterion for localization \cite{edwards72} is the sensitivity $R$ of MB energies to flux $j^s_{nn}=\mathrm{d}e_n(\varphi)/\mathrm{d}\varphi$ \cite{kohn64,castella96}. Changing the boundary conditions from PBC to antiperiodic ones \cite{edwards72} with $\delta\varphi=\pi/L$ and combined with Eq.~\eqref{dsw} gives \cite{prelovsek23}
\begin{equation}
R \equiv \frac{ \delta \varphi \; \sqrt{\overline{(d e_n(\varphi)/d \varphi)^2}} }{\Delta e} 
= \frac{ \delta \varphi \; \sqrt{ \overline{ (j^s_{nn})^2 } } }{\Delta e} \simeq \sqrt{\frac{ \delta \varphi \; \tilde{\chi}_s^0 {\cal D}_s^0 }{Y \Delta \epsilon} }\,.
\label{rrr}
\end{equation}
where ${\cal D}_s^0={\cal D}_s(\omega\to0)$ is d.c. spin diffusion constant and $\Delta e=\overline{e_{n+1}-e_n}$ is the average level spacing, or equivalently, $\rho=1/\Delta e$ the MB density of states. We note that the Thouless criterion, applied to the noninteracting Anderson model \cite{edwards72}, requires $R<1$ for localization, but also that $R$ decreases/vanishes with increasing $L$. It is also evident that in the finite-$L$ system with $R<1$, one generally cannot expect the validity of RMT universality since the MB levels do not vary enough with $\varphi$ to satisfy the concept of avoided crossing. Consequently, the level sensitivity $R\sim1$ is analogous to the criterion of merging the Heisenberg vs. Thouless time $\tau_{\mathrm H} \sim \tau_{\mathrm Th}$ \cite{suntajs20}, where
$\tau_{\mathrm H}\sim 1/\Delta e$ and $\tau_{\mathrm Th}\sim L^2/{\cal D}^0_s$.
 
\begin{figure}[!htb]
\includegraphics[width=1.0\columnwidth]{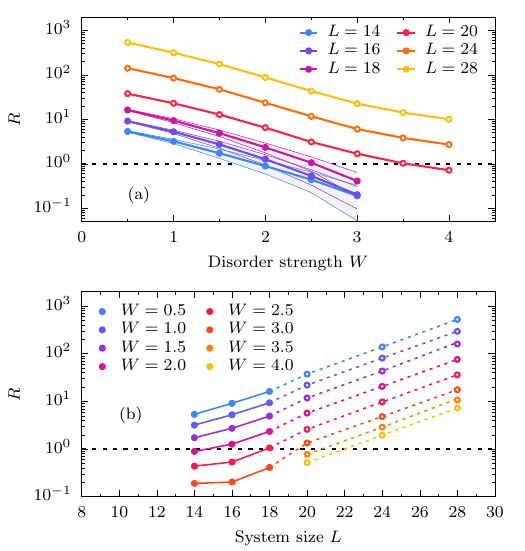}
\caption{Level sensitivity $R$: (a) vs. $W$ for different $L$: for $L=14-18$ obtained via direct ED, and for $L=20-28$ via MCLM with approximated $\Delta e$ (see text for details), (b) the same averaged $R$ vs. $L$ for different $W$. The results for $L\leq18$ ($L\geq20$) were obtained as average over $N_c=10$ ($N_c=100$) random field configurations.}
\label{fig2}
\end{figure}

In Fig.~\ref{fig1}, we present results for RMT markers for the spin current, $Y_s$, $Q_s$, and the energy current $Y_e$, $Q_e$, as well as the level sensitivity $R$, evaluated according to Eq.~\eqref{rrr}. Results are obtained via ED in the chain with $L=18$ sites for a few ($N_c=10$) fixed random configurations $\eta_l =[-1,1]$ with fields $w_l=W\eta_l$ steadily increasing with $W$. The sums over eigenstates in Eq.~\eqref{dsw} are here performed over one-quarter of states, i.e., $\tilde{N}_{st}=N_{st}/4$ states in the middle of the MB spectrum. The presented results offer some nontrivial conclusions: (a) The variation of RMT markers vs. $W$ in Figs.~\ref{fig1}(a,b) is quite similar for spin and energy current matrix elements. (b) The deviations for GOE values $Y_{s,e} \sim 1$ and $Q_{s,e} \sim 3$ appear in each sample quite simultaneously with the crossover at $R \sim 1$. (c) While the sample-sample fluctuations of d.c. values ${\cal D}^0_{s,e}$ are already substantial at intermediate $W\sim1.5-2$ (see the discussion below), each sample still reveals $Q_{s,e}\sim3$. This allows the conclusion that for fixed $w_l$ configurations, the variations of the calculated d.c. values within rather broad MB spectrum $e_n$ (one quarter of) are quite modest.

The breakdown with increasing $W$ of GOE for $Y$, $Q$ markers, as well as the marginal criterion $R(W) \sim 1$, depends on the particular sample, but even more pronounced is the dependence on the system length $L$. In Fig.~\ref{fig2}, we present the variation of sample averaged values $R$ vs. $W$ for different $L$ and vs. $L$ for different $W$. Results for ${\cal D}_s^0$ on $L=14,16,18$ chains are obtained using Eq.~\eqref{rrr} with ED and finite $\varphi>0$ ($N_c=10$ samples), while $L=20,24,28$ are evaluated via micro-canonical Lanczos method (MCLM) ($N_c=100$ samples), explained in more detail later. In the latter case, the corresponding $\Delta e$ in Eq.~\eqref{rrr} can be approximated by $L\gg1$ density of states $1/\Delta e=\rho(e)$ at $e=0$, i.e.,
\begin{equation}
\rho(e)=\frac{N_{st}}{\sqrt{2\pi}\sigma_H}\exp\left(-e^2/2\sigma_H^2\right)\to
\Delta e=\frac{\sqrt{2\pi}\sigma_H}{N_{st}}\,,
\label{deltae}
\end{equation}
where \mbox{$N_{st}=\binom{L}{L/2}$} and $\sigma_H$ can be evaluated analytically by high-$T$ moments expansion of $H$,
\begin{equation}
\sigma_H^2=\langle H^2\rangle=L\left(\frac{J^2}{16}\left(2+\Delta\right)+\frac{W^2}{32}\right)\,.
\end{equation}
We note that MCLM results at larger $W>3.5$ might already be affected by finite $\delta\omega$ resolution. Still, we find even at the largest $W\sim4$ clearly $R>1$ for the largest available $L$, as well as increasing $R(L)$. Importantly, such behavior does not conform to Thouless's criterion for localization. 

\section{Distribution of diffusion constants}

In the following, we present and analyze results for dynamical spin and energy diffusion ${\cal D}_{s,e}(\omega)$ as well as corresponding d.c. values ${\cal D}_{s,e}^0$, obtained via ED in systems with $L\leq18$, but also using upgraded microcanonical Lanczos method (MCLM) \cite{long03,karahalios09,prelovsek11,prelovsek21} with high $\delta\omega$ resolution, allowing to reach considerably larger $L\le28 $. The method (now with $\varphi=0$) evaluates dynamical current correlations, Eq.~\eqref{dsw} and Eq.~\eqref{dew}, within a microcanonical state corresponding to energy ${\cal E}$. The latter is chosen in the middle of the MB spectrum, i.e., ${\cal E}\sim0$, and with small energy dispersion $\sigma_{\cal E}<\delta\omega\sim\Delta E_\mathrm{span}/M_L$, obtained via large number of Lanczos iterations $M_L$, whereby $\Delta E_\mathrm{span}$ is the system MB energy span. In the following, we present results for $L\leq28$ sites in the $S^z_{tot}=0$ sector, with the number of MB states $N_{st}\sim10^7$ states. Using $M_L\sim5\cdot10^5$ Lanczos steps, we reach $\delta\omega\sim10^{-4}$ frequency resolution.

It has been previously numerically established, at least for the spin diffusion ${\cal D}_s^0$ \cite{herbrych22}, that results reveal a broad distribution of values in reachable systems $L \lesssim 28$. The latter cannot be captured within a simple Gaussian distribution even at intermediate $W \sim 2$, becoming closer to the log-normal distribution up to the (here numerically) reachable $W^*\lesssim4$. We note that such anomalous distributions are very distinct from the Anderson model of noninteracting particles in higher dimension $D>1$, where, due to large reachable sizes $L^D\sim10^7$ \cite{prelovsek21} evidently yield already self-averaged results. 

\begin{figure}[!htb]
\includegraphics[width=1.0\columnwidth]{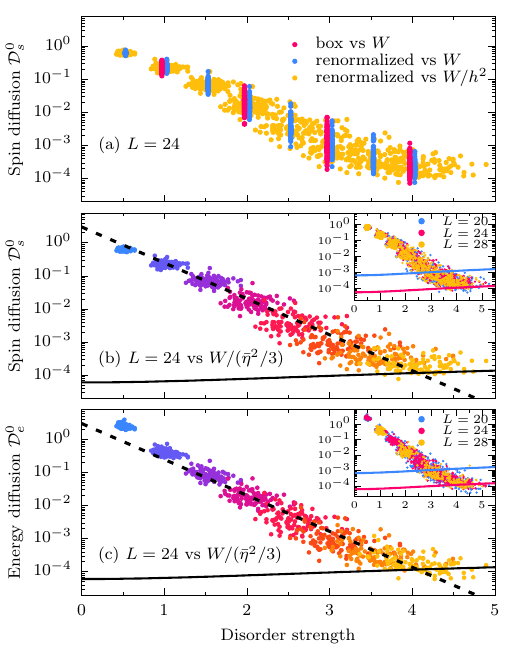}
\caption{ (a) Different presentations of the distribution of d.c. spin diffusion values ${\cal D}^0_s$: (1) vs. unrenormalized $W$, (2) using renormalized variance $\sum_l\eta^2_l/L=1/3$, and (3) presenting results vs. effective $\tilde{W}=W/(\bar{\eta}^2/3)$ (see text for details). (b) and (c) shows the distribution of results for ${\cal D}_s^0$ and ${\cal D}_e^0$, respectively, vs. effective $\tilde{W}$ obtained via MCLM for $L=24$, while the inset displays also the comparison with $L=20$ and $L=28$ results. Solid lines refer to the level sensitivity $R=1$, i.e., the estimated limit of validity of GOE. Dashed lines in panels (b,c) depict $\propto\exp(-bW/J)$ with $b=2.5$. The presented results are obtained from $N_c=100$ random-field samples.}
\label{fig3}
\end{figure}

We present in Fig.~\ref{fig3}(a) different ways of presentation of the distribution of results. (1) If we assume fixed $W$ and generate $w_l=W\eta_l$ with $\eta_l=[-1,1]$ within the box, this yields one distribution of ${\cal D}_s^0$ vs. $W$ (denoted with red color). (2) One can employ the normalized $\tilde{\eta}_l$ by requiring $\sum_l\tilde{\eta}^2_l/L=1/3$, and this simple renormalization already reduces the fluctuations of ${\cal D}^0$ by a factor $\sim2$ (denoted with blue color). (3) One can present the above "renormalized" results vs. effective $\tilde{W}=W/(\bar{\eta}^2/3) $ where $\bar{\eta}^2=(1/L)\sum_l\eta^2_l$. The latter representation is finally used in Fig.~\ref{fig3}(b,c) to display complete results obtained via MCLM for $L=24$ sites for diffusion constants ${\cal D}_s^0$ as well as ${\cal D}_e^0$, respectively. The insets also show the comparison with corresponding results for $L=20$ and $L=28$. In Figs.~\ref{fig3}(b,c) and in the insets, we also show the lines corresponding to $R=1$ [following from Eq.~\eqref{rrr} with $\Delta e$ given by Eq.~\eqref{deltae}], i.e., the expected limit of the breakdown of GOE and of the (macroscopic) relevance of our finite-size results. Our results also conclude that ${\cal D}_s^0$ and ${\cal D}_e^0$ reveal qualitatively very similar dependence on $W$.

\begin{figure}[!htb]
\includegraphics[width=1.0\columnwidth]{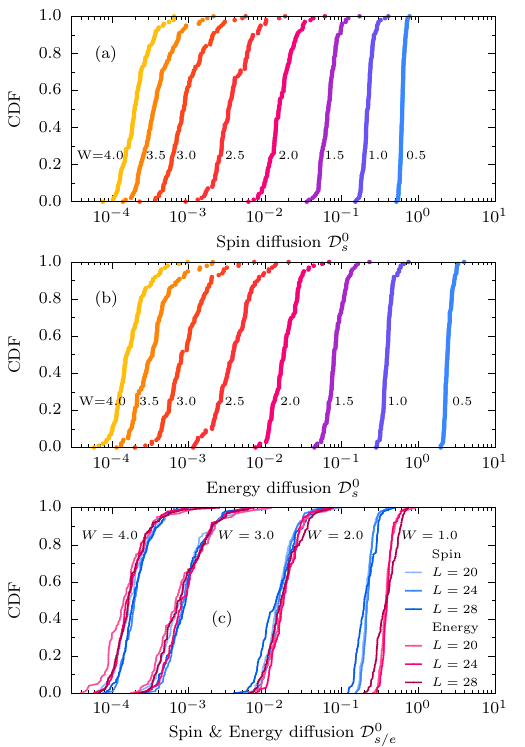}
\caption{Cumulative distribution function (CDF) for (a) d.c. spin diffusion ${\cal D}_s^0$, (b) d.c. energy diffusion ${\cal D}_e^0$, obtained via MCLM on $L=24$ sites for different $W =0.5-4$, and (c) the direct comparison between both for $W=1.0, 2.0, 3.0, 4.0$, including also $L=20\,,28$ results. The presented results are obtained from $N_c=100$ random-field samples.}
\label{fig4}
\end{figure}

In the following, we present results at renormalized $W$, i.e., with fixed $\sum_l\tilde{\eta}^2_l/L=1/3$. In Fig.~\ref{fig4}, we show one of our central results, i.e., the cumulative distribution function (CDF) for the spin and energy diffusion constants, ${\cal D}_s^0$ and ${\cal D}_e^0$, obtained via MCLM on $L=24$ sites. In the same figure, we also show a direct comparison between both of them for different system sizes $L=20\,,24\, 28$, but for restricted values of $W$. The CDF (obtained from $N_s=100$ different random field samples) shows similar qualitative shapes and the exponential-type dependence on $W$. Note also that despite the reduction due to normalization of $\tilde{\eta}$ (by a factor $\sim2$), the CDF develops from Gaussian-like at $W\lesssim2$ to much broader ones of the log-normal type at $W>2$. Finally, a direct comparison of both diffusions, ${\cal D}_s^0$ and ${\cal D}_e^0$, presented in Fig.~\ref{fig4}(c), reveals that ${\cal D}_e^0$ is considerably larger for weak $W \leq 1.5$. This agrees with the previous results \cite{mendoza19} and can be easily explained by noting that spin and energy transport are qualitatively different in the pure limit $ W \to 0$. Namely, at $W =0$, the XXZ model \eqref{hm} is integrable, and the energy transport (due to conserved $j_e$) is ballistic/dissipationless. At the same time, the spin current $j_s$ is not conserved but still superdiffusive at the considered $\Delta=1$. Consequently, these differences are also reflected in the diffusions ${\cal D}_s^0$ and ${\cal D}_e^0$ at weaker $W$. On the other hand, at larger $ W > 1.5$, both diffusions are becoming quantitatively equal. A qualitative explanation for this similarity can be found in the expression for $j_e$, Eq.~\eqref{je}, where the dominating term at large $W$ is $j_{el} \sim (w_l+w_{l+1})j_l^{l+1}/2$, which involves just local spin current. Since also in the corresponding susceptibility, the term $\tilde{\chi}_e^0 \sim W^2/12$ is dominating, both diffusions are closely related. 

\section{Evidence for subdiffusion}

\begin{figure}[!htb]
\includegraphics[width=1.0\columnwidth]{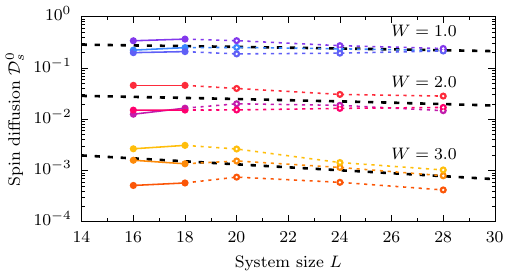}
\caption{System size $L$ dependence of spin diffusion ${\cal D}_s^0$ for $W=1,2,3$. Shown are MCLM (ED) results for $L\geq20$ ($L\leq18$) and $N_c=3$ different configurations. The dotted lines are guides to the eye, denoting the general trend with $L$.}
\label{fig5}
\end{figure}

The observed exponential-like dependence at larger \mbox{$W>1.5$} of both ${\cal D}_s^0 \sim {\cal D}_e^0\propto \mathrm{exp}(-bW/J)$ with $b\sim2.5$ indicates that we are dealing with slow, incoherent transport. It was previously shown \cite{herbrych22} that such behavior can be viewed as (essentially single-site) resonating islands satisfying \mbox{$|w_l-w_{l+d}|\lesssim\zeta$}, separated by nonresonant islands \mbox{$|w_l-w_{l+m}|>\zeta$, where $(m=1,d-1)$}. The effective exchange between resonant sites is then given by
\begin{equation}
| J_l^\mathrm{eff}| \propto \frac{ J^d} { 2^d | w_{l+1} w_{l+2} \cdots w_{l+d-1} |}
\sim \left[\frac{ J}{ 2W} \ln{\frac{W}{\zeta}} \right]^d\,.
\label{jeff}
\end{equation}
Notably, the scenario of such transport is qualitatively different from the apparently similar treatment of the hopping of noninteracting particles as in the higher-dimensional Anderson model. Namely, there the resonant condition would require $\zeta\sim J_l^\mathrm{eff} $, which cannot be satisfied self-consistently for larger $W$ \cite{prelovsek21}, leading finally to well-established localization above critical $W>W_c$. This is not the case for the MBL problem, where the resonant condition is effectively weakened by possible energy difference due to the interaction term, i.e., $\zeta\propto\Delta J\gg J^{\mathrm{eff}}_l$. This leads to the effective exchange distance $d\propto W/\zeta$ and finally ${\cal D}^0_s\propto\langle J^\mathrm{eff}_l \rangle\propto\mathrm{exp}(-\tilde{b} W/\zeta)$ where $\tilde{b}\sim\ln[(2W/J)/\ln{(W/\zeta)}]$, which for relevant $W\sim W^*\sim4$ and $\Delta=1$ comes close to observed $\tilde{b}\sim b\sim2.5$.

The above simple scenario for diffusion in MBL systems has implications. Namely, it should also apply to spin chains with quasiperiodic (QP) fields studied both theoretically/numerically \cite{iyer13,barlev17,khemani17,setiawan17,znidaric18,zhang18,aramthottil21,sierant22,prelovsek23}, as well as experimentally \cite{schreiber15,luschen17}. Indeed, similar exponential dependence on field strength $W$ has been found for ${\cal D}_s^0$ in QP systems \cite{prelovsek23}. With the essential difference that in the latter deterministic case (the absence of randomness), there are no significant (up to boundary effects) variations of ${\cal D}_s^0$ with actual field configuration. In contrast, for random but uncorrelated fields, as discussed in this study, there are large variations in $J_l^\mathrm{eff}$ resulting in variations of the effective distance between resonant islands $d$, as well as $w_l$ in the localized regions. These ingredients are the argument for the Griffiths scenario leading to domination and blocking of longer insulating islands \cite{agarwal15,gopal16,luschen16,herbrych22}. Furthermore, such arguments are also the basis for the analytical study supporting the subdiffusive transport in the thermodynamic limit \cite{deroeck24}.

\begin{figure}[!htb]
\includegraphics[width=1.0\columnwidth]{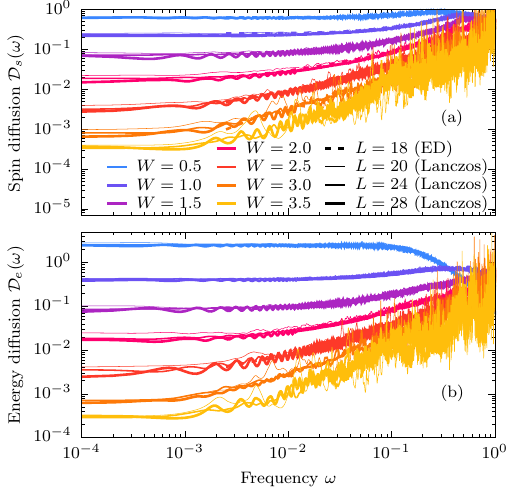}
\caption{Dynamical (a) spin and (b) energy diffusivity ${\cal D}_{s,e}(\omega)$ in the log-log scale, calculated for a single field configuration and $W=0.5-3.5$ on chains with different lengths, i.e., $L=18$ using ED, and $L=20-28$ employing MCLM for one random field sample.}
\label{fig6}
\end{figure}

One can expect the signatures of subdiffusion in the distributions of d.c. ${\cal D}^0_{s,e}$ as well as in the dynamical response ${\cal D}_{s,e}(\omega)$. Previous consideration of the toy model \cite{herbrych22} hints that Griffiths effect of large insulating islands should show up in the enhanced low-diffusion tails in CDF and slow reduction of the average and typical d.c. diffusion values with increasing $L$. Still, even the toy model reveals that very large sizes, $L\gg100$, are needed to establish this trend clearly. In our studies, we have a rather small span, $14\leq L\leq28$. Hence, the variation of ${\cal D}^0_{s,e}$ with $L$, as presented in Fig.~\ref{fig5} for a few field configurations and chosen $W$, do not reveal a pronounced $L$ dependence for smaller $W \le 2.0$, but apparently more visible reduction with $L$ for larger $W \geq 3.0$. 

In contrast, one expects more pronounced effects in the dynamical diffusivity ${\cal D}_{s,e}(\omega)$ since results cover a much larger dynamical range, taking into account also $\omega > \delta \omega \sim 10^{-4}$. As the signature of subdiffusion one can identify the dependence ${\cal D}(\omega) \sim {\cal D}^0+c|\omega|^\alpha$ with $\alpha<1$ \cite{agarwal15,gopal16} in the low \mbox{$\omega<\omega^*$} range. Such behavior implies that even with vanishing (or very small) ${\cal D}^0$, the d.c. polarizability \mbox{$\chi_p\propto\int\mathrm{d}\,\omega{\cal D}_s(\omega)/\omega^2$} \cite{prelovsek17} is diverging, in contrast to the localization defined by $\chi_p<\infty$ requiring $\alpha>1$. Numerical investigation of finite systems indicated that ${\cal D}_{s,e}(\omega)$ [and related conductivity $\sigma_s(\omega)$] at $\omega\sim {\cal O}(0.1)$ is indeed consistent with $\alpha\simeq 1$ \cite{karahalios09,barisic10,gopal15,steinigeweg16}. A similar conclusion (in a similar frequency range) can be reached by analysis of the short wave-length $q\to\pi$ response (i.e., the so-called imbalance) \cite{mierzejewski16,prelovsek17a,herbrych22}. Here, contrary, we show that a closer analysis of $\omega\ll1$ (obtained with upgraded MCLM method with $\delta\omega\sim10^{-4}$ resolution) indicates that random-field results are generally at larger $W$ consistent with $\alpha<1$. The latter contrasts the QP field results, for which $\alpha\simeq 1$ is reported down to $\omega\to0$.

\begin{figure}[htb!]
\includegraphics[width=1.0\columnwidth]{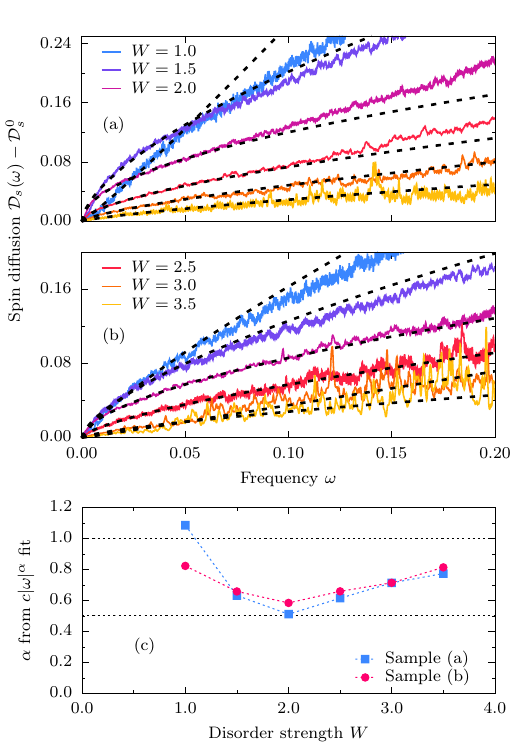}
\caption{(a,b) Dynamical spin diffusivity ${\cal D}_s(\omega)-{\cal D}_s^0$ for two random field configurations with few $W$, calculated on $L=28$ chain. Each panel represents a different $w_l$ configuration. The black dashed lines correspond to the fits to the $c|\omega|^\alpha$ function in the $0<\omega<0.05$ region. Panel (c) depicts $\alpha$ dependence on the strength of the disorder $W$ (obtained from the fit).}
\label{fig7}
\end{figure}

\begin{figure}[htb!]
\includegraphics[width=1.0\columnwidth]{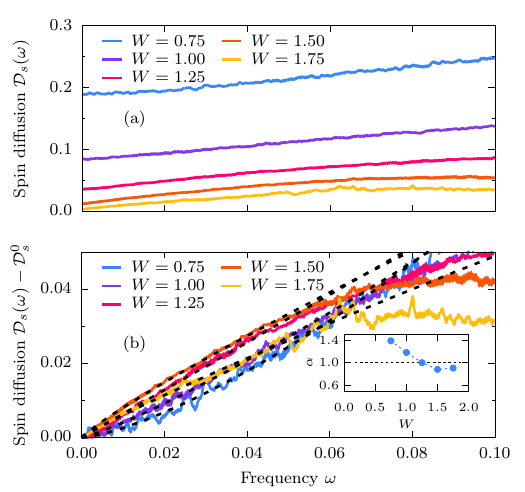}
\caption{(a) Dynamical spin diffusivity ${\cal D}_s(\omega)$, calculated via MCLM for for quasiperiodic system ($L=27$). See text for details. (b) Frequency dependence of ${\cal D}_s(\omega)-{\cal D}_s^0$, together with fits to the $c|\omega|^\alpha$ function (black, dashed lines). The inset depicts $\alpha$ dependence on the (quasiperiodic) disorder strength.}
\label{fig8}
\end{figure}

In Fig.~\ref{fig6}, we present numerical results for dynamical ${\cal D}_{s,e}(\omega)$, obtained for single field configurations but different $W=0.5-3.5$ on different system sizes $L=18-28$. While $L =18$ can be reached by ED directly evaluating Eq.~\eqref{dsw}, $L\geq20$ results are obtained via MCLM with finite frequency resolution $\delta\omega\sim 10^{-4}$. Apart from fluctuations in MCLM results, being the consequence of finite $\delta\omega$ as well as the restricted sampling over initial microcanonical energies ${\cal E}$, results are consistent, provided that we input the same field configurations $w_l$. The dependence of MCLM results on specific $w_l$ configurations is presented in Fig.~\ref{fig7} for two different field configurations [with samples chosen from the middle of the distribution presented in Fig.~\ref{fig4}(a)]. To focus on the frequency dependence we present ${\cal D}_{s}(\omega)-{\cal D}^0_{s}$, which allows for the analysis of $\propto|\omega|^\alpha$ fit in the $0<\omega<0.05$ region. It is evident from the results presented in Fig.~\ref{fig7}(c), that $\alpha<1$ for $W\gtrsim 1$ and $\alpha\simeq0.5$ for $W\simeq 2$ for both considered samples. Furthermore, our results indicate that although $\alpha<1$ for all considered $1<W<4$, its value slowly approaches $\alpha\sim 1$ for increasing $W$. Such linear frequency dependence of ${\cal D}_{s}(\omega)$ was previously reported \cite{agarwal15,gopal15,potter15,mierzejewski16,steinigeweg16,barisic16} in proximity to the (presumable) transition to MBL phase, $W>W_c$, based on the quantities averaged over random field configurations. Note that the true MBL phase reacquires ${\cal D}_s^0=0$ as well as $\alpha>1$. Furthermore, there are also some numerical evidence that characteristic subdiffusion scale $\omega^*$ is shrinking with disorder strength $W$, e.g., looking at Fig.~\ref{fig7}a. The argument could be that $\omega^*$ should decrease together with strongly decreasing ${\cal D}_s^0$. Still, evidence is not conclusive, mostly due to numerical restrictions due to finite resolution $\delta \omega$ at large $W$.

Finally, we comment on the difference between the random-field and QP model. Note that the crossover to the MBL phase in the latter was found for $W\simeq1.5$ \cite{setiawan17,prelovsek23}. The QP field configuration is obtained with $w_l=W\cos(ql+\phi_0)$, (nearly) irrational $q/(2\pi)=11/27\sim(3-\sqrt{5})/2$, and $\phi_0=0$. In Fig.~\ref{fig8}(a), we show the full ${\cal D}_s(\omega)$ MCLM spectra (for $L=27$ sites), while in panel (b) we again focus on the frequency dependence of ${\cal D}_{s}(\omega)-{\cal D}^0_{s}$ with fits to $|\omega|^\alpha$. It is evident that the results for QP at low-$\omega$ are different: for all considered values $W$ we observe $\alpha\gtrsim 1$.

\section{Summary} 

Besides the more frequently discussed spin diffusion, we investigate in this paper the energy diffusion within the XXZ Heisenberg spin chain, which is subject to random local fields. The memory-function approach in the hydrodynamic regime allows the derivation of appropriate Einstein relations, expressing diffusion response in terms of spin and energy current correlations, respectively. Following numerical results on finite systems with up to $L=28$ sites for both dynamical diffusivities ${\cal D}_{s,e}(\omega)$ as well as for d.c. diffusion constants ${\cal D}^0_{s,e}$ reveal some essential similarities but as well differences. Static ${\cal D}^0_{s,e}$ shows strong exponential-like dependence on the field strength $W$ but also pronounced dependence and variation with the actual random field configurations $w_l$. While energy diffusion is evidently faster, i.e. ${\cal D}^0_e>{\cal D}^0_s $ at weak $W<1.5$ \cite{mendoza19}, both diffusion become (up to numerical resolution) nearly equal at larger $W>2.0$, both governed by the incoherent spin transport. This conclusion is important in relation to the frequently discussed Thouless time determined by the slowest diffusion $\tau_\mathrm{Th}=L^2/{\cal D}^0$, where our results indicate that at large disorders, both diffusions are equally important.

The Thouless approach relating the particle/spin transport to the sensitivity $R$ to the imposed flux (or twisted boundary conditions) allows, besides the direct test of (Thouless) localization criteria, also the closer investigation of whether the finite-size systems comply with the ETH and RMT universality expected for normal ergodic systems. Here, it should be stressed that even at intermediate $W$, where transport properties reveal considerable variations with actual field configurations, the properties of current matrix elements within individual samples satisfy well the criteria of GOE universality. In our finite systems, the breakdown of GOE and RMT can be directly related to crossover in the level sensitivity $R\sim1$. On the other hand, at the same disorder, the observed systematic growth $R$ with $L$ gives a clear message that this is still not localization (MBL), at least not within the (numerically) reachable regime $W\lesssim4$.

Due to modest system sizes $L\leq28$, the values of the d.c. diffusion constants ${\cal D}^0_{s,e}$ show large fluctuations between different field configurations at the same effective strength $W$. We can partly reduce these variations (at the same $W$) by fixing/renormalizing the random amplitudes $\sum_l\tilde{\eta}^2_l=1/3$. Still, at larger $W$, the distributions of ${\cal D}^0_{s,e}$ are becoming non-Gaussian, i.e., closer to a log-normal distribution, even with pronounced tails. 

The behavior at large $W$ offers the analogy with the Griffiths scenario of large transport fluctuations due to small resonant islands separated by long (and sample-dependent) nonresonant regions. This possibility can be tested by comparing analogous results for the model with quasiperiodic fields, where, due to deterministic potential, the Griffiths scenario is not expected. Indeed, by evaluating and analyzing the dynamical diffusivity ${\cal D}_s(\omega)\sim{\cal D}_s^0+c|\omega|^\alpha $ for QP fields we get at larger $W$ universally $\alpha \sim 1$ coexisting with nonzero, but exponentially decaying dc ${\cal D}_s^0(W)$ \cite{prelovsek23}. In clear contrast, random fields typically reveal $\alpha<1$, characteristic for subdiffusion \cite{deroeck24} and the Griffiths effects \cite{agarwal15,gopal16}. Still, the latter scenario also requires diminishing ${\cal D}_s^0$ with $L$, which remains a hard numerical challenge, and we only marginally confirm it.
 
\noindent{\it Acknowledgments.}
P.P. acknowledges the support of the N1-0088 project of the Slovenian Research Agency. The numerical calculations were partly carried out at the facilities of the Wroclaw Centre for Networking and Supercomputing.\\
The data that support the findings of this article are openly available \cite{opendata}.

\bibliography{manudifen}

\begin{thebibliography}{67}%
\makeatletter
\providecommand \@ifxundefined [1]{%
 \@ifx{#1\undefined}
}%
\providecommand \@ifnum [1]{%
 \ifnum #1\expandafter \@firstoftwo
 \else \expandafter \@secondoftwo
 \fi
}%
\providecommand \@ifx [1]{%
 \ifx #1\expandafter \@firstoftwo
 \else \expandafter \@secondoftwo
 \fi
}%
\providecommand \natexlab [1]{#1}%
\providecommand \enquote  [1]{``#1''}%
\providecommand \bibnamefont  [1]{#1}%
\providecommand \bibfnamefont [1]{#1}%
\providecommand \citenamefont [1]{#1}%
\providecommand \href@noop [0]{\@secondoftwo}%
\providecommand \href [0]{\begingroup \@sanitize@url \@href}%
\providecommand \@href[1]{\@@startlink{#1}\@@href}%
\providecommand \@@href[1]{\endgroup#1\@@endlink}%
\providecommand \@sanitize@url [0]{\catcode `\\12\catcode `\$12\catcode
  `\&12\catcode `\#12\catcode `\^12\catcode `\_12\catcode `\%12\relax}%
\providecommand \@@startlink[1]{}%
\providecommand \@@endlink[0]{}%
\providecommand \url  [0]{\begingroup\@sanitize@url \@url }%
\providecommand \@url [1]{\endgroup\@href {#1}{\urlprefix }}%
\providecommand \urlprefix  [0]{URL }%
\providecommand \Eprint [0]{\href }%
\providecommand \doibase [0]{https://doi.org/}%
\providecommand \selectlanguage [0]{\@gobble}%
\providecommand \bibinfo  [0]{\@secondoftwo}%
\providecommand \bibfield  [0]{\@secondoftwo}%
\providecommand \translation [1]{[#1]}%
\providecommand \BibitemOpen [0]{}%
\providecommand \bibitemStop [0]{}%
\providecommand \bibitemNoStop [0]{.\EOS\space}%
\providecommand \EOS [0]{\spacefactor3000\relax}%
\providecommand \BibitemShut  [1]{\csname bibitem#1\endcsname}%
\let\auto@bib@innerbib\@empty
\bibitem [{\citenamefont {Anderson}(1958)}]{anderson58}%
  \BibitemOpen
  \bibfield  {author} {\bibinfo {author} {\bibfnamefont {P.~W.}\ \bibnamefont
  {Anderson}},\ }\bibfield  {title} {\bibinfo {title} {{Absence of Diffusion in
  Certain Random Lattices}},\ }\href {https://doi.org/10.1103/PhysRev.109.1492}
  {\bibfield  {journal} {\bibinfo  {journal} {Phys. Rev.}\ }\textbf {\bibinfo
  {volume} {109}},\ \bibinfo {pages} {1492} (\bibinfo {year}
  {1958})}\BibitemShut {NoStop}%
\bibitem [{\citenamefont {Abrahams}\ \emph {et~al.}(1979)\citenamefont
  {Abrahams}, \citenamefont {Anderson}, \citenamefont {Licciardello},\ and\
  \citenamefont {Ramakrishnan}}]{abrahams79}%
  \BibitemOpen
  \bibfield  {author} {\bibinfo {author} {\bibfnamefont {E.}~\bibnamefont
  {Abrahams}}, \bibinfo {author} {\bibfnamefont {P.~W.}\ \bibnamefont
  {Anderson}}, \bibinfo {author} {\bibfnamefont {D.~C.}\ \bibnamefont
  {Licciardello}},\ and\ \bibinfo {author} {\bibfnamefont {T.~V.}\ \bibnamefont
  {Ramakrishnan}},\ }\bibfield  {title} {\bibinfo {title} {{Scaling theory of
  localization: absense of quantum diffusion in two dimensions}},\ }\href
  {https://doi.org/10.1103/PhysRevLett.42.673} {\bibfield  {journal} {\bibinfo
  {journal} {Phys. Rev. Lett.}\ }\textbf {\bibinfo {volume} {42}},\ \bibinfo
  {pages} {673} (\bibinfo {year} {1979})}\BibitemShut {NoStop}%
\bibitem [{\citenamefont {Evers}\ and\ \citenamefont {Mirlin}(2008)}]{evers08}%
  \BibitemOpen
  \bibfield  {author} {\bibinfo {author} {\bibfnamefont {F.}~\bibnamefont
  {Evers}}\ and\ \bibinfo {author} {\bibfnamefont {A.~D.}\ \bibnamefont
  {Mirlin}},\ }\bibfield  {title} {\bibinfo {title} {{Anderson transitions}},\
  }\href {https://doi.org/10.1103/RevModPhys.80.1355} {\bibfield  {journal}
  {\bibinfo  {journal} {Rev. Mod. Phys.}\ }\textbf {\bibinfo {volume} {80}},\
  \bibinfo {pages} {1355} (\bibinfo {year} {2008})}\BibitemShut {NoStop}%
\bibitem [{\citenamefont {Kramer}\ and\ \citenamefont
  {MacKinnon}(1993)}]{kramer93}%
  \BibitemOpen
  \bibfield  {author} {\bibinfo {author} {\bibfnamefont {B.}~\bibnamefont
  {Kramer}}\ and\ \bibinfo {author} {\bibfnamefont {A.}~\bibnamefont
  {MacKinnon}},\ }\bibfield  {title} {\bibinfo {title} {{Localization: theory
  and experiment}},\ }\href {https://doi.org/10.1088/0034-4885/56/12/001}
  {\bibfield  {journal} {\bibinfo  {journal} {Rep. Prog. Phys.}\ }\textbf
  {\bibinfo {volume} {56}},\ \bibinfo {pages} {1469} (\bibinfo {year}
  {1993})}\BibitemShut {NoStop}%
\bibitem [{\citenamefont {Markos}(2006)}]{markos06}%
  \BibitemOpen
  \bibfield  {author} {\bibinfo {author} {\bibfnamefont {P.}~\bibnamefont
  {Markos}},\ }\bibfield  {title} {\bibinfo {title} {{Numerical analysis of the
  anderson localization}},\ }\href {https://doi.org/10.2478/v10155-010-0081-0}
  {\bibfield  {journal} {\bibinfo  {journal} {Acta Phys. Slovaca}\ }\textbf
  {\bibinfo {volume} {56}},\ \bibinfo {pages} {561} (\bibinfo {year}
  {2006})}\BibitemShut {NoStop}%
\bibitem [{\citenamefont {Basko}\ \emph {et~al.}(2006)\citenamefont {Basko},
  \citenamefont {Aleiner},\ and\ \citenamefont {Altshuler}}]{basko06}%
  \BibitemOpen
  \bibfield  {author} {\bibinfo {author} {\bibfnamefont {D.}~\bibnamefont
  {Basko}}, \bibinfo {author} {\bibfnamefont {I.}~\bibnamefont {Aleiner}},\
  and\ \bibinfo {author} {\bibfnamefont {B.}~\bibnamefont {Altshuler}},\
  }\bibfield  {title} {\bibinfo {title} {{Metal--insulator transition in a
  weakly interacting many-electron system with localized single-particle
  states}},\ }\href {https://doi.org/10.1016/j.aop.2005.11.014} {\bibfield
  {journal} {\bibinfo  {journal} {Ann. Phys. (N. Y.)}\ }\textbf {\bibinfo
  {volume} {321}},\ \bibinfo {pages} {1126} (\bibinfo {year}
  {2006})}\BibitemShut {NoStop}%
\bibitem [{\citenamefont {Oganesyan}\ and\ \citenamefont
  {Huse}(2007)}]{oganesyan07}%
  \BibitemOpen
  \bibfield  {author} {\bibinfo {author} {\bibfnamefont {V.}~\bibnamefont
  {Oganesyan}}\ and\ \bibinfo {author} {\bibfnamefont {D.~A.}\ \bibnamefont
  {Huse}},\ }\bibfield  {title} {\bibinfo {title} {{Localization of interacting
  fermions at high temperature}},\ }\href
  {https://doi.org/10.1103/PhysRevB.75.155111} {\bibfield  {journal} {\bibinfo
  {journal} {Phys. Rev. B}\ }\textbf {\bibinfo {volume} {75}},\ \bibinfo
  {pages} {155111} (\bibinfo {year} {2007})}\BibitemShut {NoStop}%
\bibitem [{\citenamefont {Luitz}\ \emph {et~al.}(2015)\citenamefont {Luitz},
  \citenamefont {Laflorencie},\ and\ \citenamefont {Alet}}]{luitz15}%
  \BibitemOpen
  \bibfield  {author} {\bibinfo {author} {\bibfnamefont {D.~J.}\ \bibnamefont
  {Luitz}}, \bibinfo {author} {\bibfnamefont {N.}~\bibnamefont {Laflorencie}},\
  and\ \bibinfo {author} {\bibfnamefont {F.}~\bibnamefont {Alet}},\ }\bibfield
  {title} {\bibinfo {title} {{Many-body localization edge in the random-field
  {Heisenberg} chain}},\ }\href {https://doi.org/10.1103/PhysRevB.91.081103}
  {\bibfield  {journal} {\bibinfo  {journal} {Phys. Rev. B}\ }\textbf {\bibinfo
  {volume} {91}},\ \bibinfo {pages} {081103} (\bibinfo {year}
  {2015})}\BibitemShut {NoStop}%
\bibitem [{\citenamefont {Serbyn}\ and\ \citenamefont
  {Moore}(2016)}]{serbyn16}%
  \BibitemOpen
  \bibfield  {author} {\bibinfo {author} {\bibfnamefont {M.}~\bibnamefont
  {Serbyn}}\ and\ \bibinfo {author} {\bibfnamefont {J.~E.}\ \bibnamefont
  {Moore}},\ }\bibfield  {title} {\bibinfo {title} {{Spectral statistics across
  the many-body localization transition}},\ }\href
  {https://doi.org/10.1103/PhysRevB.93.041424} {\bibfield  {journal} {\bibinfo
  {journal} {Phys. Rev. B}\ }\textbf {\bibinfo {volume} {93}},\ \bibinfo
  {pages} {041424} (\bibinfo {year} {2016})}\BibitemShut {NoStop}%
\bibitem [{\citenamefont {\v{S}untajs}\ \emph
  {et~al.}(2020{\natexlab{a}})\citenamefont {\v{S}untajs}, \citenamefont
  {Bon\v{c}a}, \citenamefont {Prosen},\ and\ \citenamefont
  {Vidmar}}]{suntajs20}%
  \BibitemOpen
  \bibfield  {author} {\bibinfo {author} {\bibfnamefont {J.}~\bibnamefont
  {\v{S}untajs}}, \bibinfo {author} {\bibfnamefont {J.}~\bibnamefont
  {Bon\v{c}a}}, \bibinfo {author} {\bibfnamefont {T.}~\bibnamefont {Prosen}},\
  and\ \bibinfo {author} {\bibfnamefont {L.}~\bibnamefont {Vidmar}},\
  }\bibfield  {title} {\bibinfo {title} {{Quantum chaos challenges many-body
  localization}},\ }\href {https://doi.org/10.1103/PhysRevE.102.062144}
  {\bibfield  {journal} {\bibinfo  {journal} {Phys. Rev. E}\ }\textbf {\bibinfo
  {volume} {102}},\ \bibinfo {pages} {062144} (\bibinfo {year}
  {2020}{\natexlab{a}})}\BibitemShut {NoStop}%
\bibitem [{\citenamefont {Sierant}\ \emph {et~al.}(2020)\citenamefont
  {Sierant}, \citenamefont {Delande},\ and\ \citenamefont
  {Zakrzewski}}]{sierant20}%
  \BibitemOpen
  \bibfield  {author} {\bibinfo {author} {\bibfnamefont {P.}~\bibnamefont
  {Sierant}}, \bibinfo {author} {\bibfnamefont {D.}~\bibnamefont {Delande}},\
  and\ \bibinfo {author} {\bibfnamefont {J.}~\bibnamefont {Zakrzewski}},\
  }\bibfield  {title} {\bibinfo {title} {{Thouless Time Analysis of Anderson
  and Many-Body Localization Transitions}},\ }\href
  {https://doi.org/10.1103/PhysRevLett.124.186601} {\bibfield  {journal}
  {\bibinfo  {journal} {Phys. Rev. Lett.}\ }\textbf {\bibinfo {volume} {124}},\
  \bibinfo {pages} {186601} (\bibinfo {year} {2020})}\BibitemShut {NoStop}%
\bibitem [{\citenamefont {\v{Z}nidari\v{c}}\ \emph {et~al.}(2008)\citenamefont
  {\v{Z}nidari\v{c}}, \citenamefont {Prosen},\ and\ \citenamefont
  {Prelov\v{s}ek}}]{znidaric08}%
  \BibitemOpen
  \bibfield  {author} {\bibinfo {author} {\bibfnamefont {M.}~\bibnamefont
  {\v{Z}nidari\v{c}}}, \bibinfo {author} {\bibfnamefont {T.}~\bibnamefont
  {Prosen}},\ and\ \bibinfo {author} {\bibfnamefont {P.}~\bibnamefont
  {Prelov\v{s}ek}},\ }\bibfield  {title} {\bibinfo {title} {{Many-body
  localization in the Heisenberg {$XXZ$} magnet in a random field}},\ }\href
  {https://doi.org/10.1103/PhysRevB.77.064426} {\bibfield  {journal} {\bibinfo
  {journal} {Phys. Rev. B}\ }\textbf {\bibinfo {volume} {77}},\ \bibinfo
  {pages} {064426} (\bibinfo {year} {2008})}\BibitemShut {NoStop}%
\bibitem [{\citenamefont {Bardarson}\ \emph {et~al.}(2012)\citenamefont
  {Bardarson}, \citenamefont {Pollmann},\ and\ \citenamefont
  {Moore}}]{bardarson12}%
  \BibitemOpen
  \bibfield  {author} {\bibinfo {author} {\bibfnamefont {J.~H.}\ \bibnamefont
  {Bardarson}}, \bibinfo {author} {\bibfnamefont {F.}~\bibnamefont
  {Pollmann}},\ and\ \bibinfo {author} {\bibfnamefont {J.~E.}\ \bibnamefont
  {Moore}},\ }\bibfield  {title} {\bibinfo {title} {{Unbounded Growth of
  Entanglement in Models of Many-Body Localization}},\ }\href
  {https://doi.org/10.1103/PhysRevLett.109.017202} {\bibfield  {journal}
  {\bibinfo  {journal} {Phys. Rev. Lett.}\ }\textbf {\bibinfo {volume} {109}},\
  \bibinfo {pages} {017202} (\bibinfo {year} {2012})}\BibitemShut {NoStop}%
\bibitem [{\citenamefont {Serbyn}\ \emph {et~al.}(2015)\citenamefont {Serbyn},
  \citenamefont {Papi\'{c}},\ and\ \citenamefont {Abanin}}]{serbyn15}%
  \BibitemOpen
  \bibfield  {author} {\bibinfo {author} {\bibfnamefont {M.}~\bibnamefont
  {Serbyn}}, \bibinfo {author} {\bibfnamefont {Z.}~\bibnamefont {Papi\'{c}}},\
  and\ \bibinfo {author} {\bibfnamefont {D.~A.}\ \bibnamefont {Abanin}},\
  }\bibfield  {title} {\bibinfo {title} {{Criterion for Many-Body
  Localization-Delocalization Phase Transition}},\ }\href
  {https://doi.org/10.1103/PhysRevX.5.041047} {\bibfield  {journal} {\bibinfo
  {journal} {Phys. Rev. X}\ }\textbf {\bibinfo {volume} {5}},\ \bibinfo {pages}
  {041047} (\bibinfo {year} {2015})}\BibitemShut {NoStop}%
\bibitem [{\citenamefont {Berkelbach}\ and\ \citenamefont
  {Reichman}(2010)}]{berkelbach10}%
  \BibitemOpen
  \bibfield  {author} {\bibinfo {author} {\bibfnamefont {T.~C.}\ \bibnamefont
  {Berkelbach}}\ and\ \bibinfo {author} {\bibfnamefont {D.~R.}\ \bibnamefont
  {Reichman}},\ }\bibfield  {title} {\bibinfo {title} {{Conductivity of
  disordered quantum lattice models at infinite temperature :}},\ }\href
  {https://doi.org/10.1103/PhysRevB.81.224429} {\bibfield  {journal} {\bibinfo
  {journal} {Phys. Rev. B}\ }\textbf {\bibinfo {volume} {81}},\ \bibinfo
  {pages} {224429} (\bibinfo {year} {2010})}\BibitemShut {NoStop}%
\bibitem [{\citenamefont {Bari\v{s}i\'{c}}\ and\ \citenamefont
  {Prelov\v{s}ek}(2010)}]{barisic10}%
  \BibitemOpen
  \bibfield  {author} {\bibinfo {author} {\bibfnamefont {O.~S.}\ \bibnamefont
  {Bari\v{s}i\'{c}}}\ and\ \bibinfo {author} {\bibfnamefont {P.}~\bibnamefont
  {Prelov\v{s}ek}},\ }\bibfield  {title} {\bibinfo {title} {{Conductivity in a
  disordered one-dimensional system of interacting fermions}},\ }\href
  {https://doi.org/10.1103/PhysRevB.82.161106} {\bibfield  {journal} {\bibinfo
  {journal} {Phys. Rev. B}\ }\textbf {\bibinfo {volume} {82}},\ \bibinfo
  {pages} {161106} (\bibinfo {year} {2010})}\BibitemShut {NoStop}%
\bibitem [{\citenamefont {Agarwal}\ \emph {et~al.}(2015)\citenamefont
  {Agarwal}, \citenamefont {Gopalakrishnan}, \citenamefont {Knap},
  \citenamefont {M\"{u}ller},\ and\ \citenamefont {Demler}}]{agarwal15}%
  \BibitemOpen
  \bibfield  {author} {\bibinfo {author} {\bibfnamefont {K.}~\bibnamefont
  {Agarwal}}, \bibinfo {author} {\bibfnamefont {S.}~\bibnamefont
  {Gopalakrishnan}}, \bibinfo {author} {\bibfnamefont {M.}~\bibnamefont
  {Knap}}, \bibinfo {author} {\bibfnamefont {M.}~\bibnamefont {M\"{u}ller}},\
  and\ \bibinfo {author} {\bibfnamefont {E.}~\bibnamefont {Demler}},\
  }\bibfield  {title} {\bibinfo {title} {{Anomalous Diffusion and Griffiths
  Effects Near the Many-Body Localization Transition}},\ }\href
  {https://doi.org/10.1103/PhysRevLett.114.160401} {\bibfield  {journal}
  {\bibinfo  {journal} {Phys. Rev. Lett.}\ }\textbf {\bibinfo {volume} {114}},\
  \bibinfo {pages} {160401} (\bibinfo {year} {2015})}\BibitemShut {NoStop}%
\bibitem [{\citenamefont {Bar~Lev}\ \emph {et~al.}(2015)\citenamefont
  {Bar~Lev}, \citenamefont {Cohen},\ and\ \citenamefont {Reichman}}]{barlev15}%
  \BibitemOpen
  \bibfield  {author} {\bibinfo {author} {\bibfnamefont {Y.}~\bibnamefont
  {Bar~Lev}}, \bibinfo {author} {\bibfnamefont {G.}~\bibnamefont {Cohen}},\
  and\ \bibinfo {author} {\bibfnamefont {D.~R.}\ \bibnamefont {Reichman}},\
  }\bibfield  {title} {\bibinfo {title} {{Absence of Diffusion in an
  Interacting System of Spinless Fermions on a One-Dimensional Disordered
  Lattice}},\ }\href {https://doi.org/10.1103/PhysRevLett.114.100601}
  {\bibfield  {journal} {\bibinfo  {journal} {Phys. Rev. Lett.}\ }\textbf
  {\bibinfo {volume} {114}},\ \bibinfo {pages} {100601} (\bibinfo {year}
  {2015})}\BibitemShut {NoStop}%
\bibitem [{\citenamefont {Steinigeweg}\ \emph {et~al.}(2016)\citenamefont
  {Steinigeweg}, \citenamefont {Herbrych}, \citenamefont {Pollmann},\ and\
  \citenamefont {Brenig}}]{steinigeweg16}%
  \BibitemOpen
  \bibfield  {author} {\bibinfo {author} {\bibfnamefont {R.}~\bibnamefont
  {Steinigeweg}}, \bibinfo {author} {\bibfnamefont {J.}~\bibnamefont
  {Herbrych}}, \bibinfo {author} {\bibfnamefont {F.}~\bibnamefont {Pollmann}},\
  and\ \bibinfo {author} {\bibfnamefont {W.}~\bibnamefont {Brenig}},\
  }\bibfield  {title} {\bibinfo {title} {{Scaling of the Optical Conductivity
  in the Transition from Thermal to Many-Body Localized Phases}},\ }\href
  {https://doi.org/10.1103/PhysRevB.94.180401} {\bibfield  {journal} {\bibinfo
  {journal} {Phys. Rev. B}\ }\textbf {\bibinfo {volume} {94}},\ \bibinfo
  {pages} {180401(R)} (\bibinfo {year} {2016})}\BibitemShut {NoStop}%
\bibitem [{\citenamefont {Prelov\v{s}ek}\ \emph {et~al.}(2017)\citenamefont
  {Prelov\v{s}ek}, \citenamefont {Mierzejewski}, \citenamefont
  {Bari\v{s}i\'{c}},\ and\ \citenamefont {Herbrych}}]{prelovsek17}%
  \BibitemOpen
  \bibfield  {author} {\bibinfo {author} {\bibfnamefont {P.}~\bibnamefont
  {Prelov\v{s}ek}}, \bibinfo {author} {\bibfnamefont {M.}~\bibnamefont
  {Mierzejewski}}, \bibinfo {author} {\bibfnamefont {O.}~\bibnamefont
  {Bari\v{s}i\'{c}}},\ and\ \bibinfo {author} {\bibfnamefont {J.}~\bibnamefont
  {Herbrych}},\ }\bibfield  {title} {\bibinfo {title} {{Density correlations
  and transport in models of many-body localization}},\ }\href
  {https://doi.org/10.1002/andp.201600362} {\bibfield  {journal} {\bibinfo
  {journal} {Ann. Phys. (Berl.)}\ }\textbf {\bibinfo {volume} {529}},\ \bibinfo
  {pages} {1600362} (\bibinfo {year} {2017})}\BibitemShut {NoStop}%
\bibitem [{\citenamefont {Pal}\ and\ \citenamefont {Huse}(2010)}]{pal10}%
  \BibitemOpen
  \bibfield  {author} {\bibinfo {author} {\bibfnamefont {A.}~\bibnamefont
  {Pal}}\ and\ \bibinfo {author} {\bibfnamefont {D.~A.}\ \bibnamefont {Huse}},\
  }\bibfield  {title} {\bibinfo {title} {{Many-body localization phase
  transition}},\ }\href {https://doi.org/10.1103/PhysRevB.82.174411} {\bibfield
   {journal} {\bibinfo  {journal} {Phys. Rev. B}\ }\textbf {\bibinfo {volume}
  {82}},\ \bibinfo {pages} {174411} (\bibinfo {year} {2010})}\BibitemShut
  {NoStop}%
\bibitem [{\citenamefont {Serbyn}\ \emph {et~al.}(2013)\citenamefont {Serbyn},
  \citenamefont {Papi\'{c}},\ and\ \citenamefont {Abanin}}]{serbyn13}%
  \BibitemOpen
  \bibfield  {author} {\bibinfo {author} {\bibfnamefont {M.}~\bibnamefont
  {Serbyn}}, \bibinfo {author} {\bibfnamefont {Z.}~\bibnamefont {Papi\'{c}}},\
  and\ \bibinfo {author} {\bibfnamefont {D.~A.}\ \bibnamefont {Abanin}},\
  }\bibfield  {title} {\bibinfo {title} {{Local Conservation Laws and the
  Structure of the Many-Body Localized States}},\ }\href
  {https://doi.org/10.1103/PhysRevLett.111.127201} {\bibfield  {journal}
  {\bibinfo  {journal} {Phys. Rev. Lett.}\ }\textbf {\bibinfo {volume} {111}},\
  \bibinfo {pages} {127201} (\bibinfo {year} {2013})}\BibitemShut {NoStop}%
\bibitem [{\citenamefont {Huse}\ \emph {et~al.}(2014)\citenamefont {Huse},
  \citenamefont {Nandkishore},\ and\ \citenamefont {Oganesyan}}]{huse14}%
  \BibitemOpen
  \bibfield  {author} {\bibinfo {author} {\bibfnamefont {D.~A.}\ \bibnamefont
  {Huse}}, \bibinfo {author} {\bibfnamefont {R.}~\bibnamefont {Nandkishore}},\
  and\ \bibinfo {author} {\bibfnamefont {V.}~\bibnamefont {Oganesyan}},\
  }\bibfield  {title} {\bibinfo {title} {{Phenomenology of fully
  many-body-localized systems}},\ }\href
  {https://doi.org/10.1103/PhysRevB.90.174202} {\bibfield  {journal} {\bibinfo
  {journal} {Phys. Rev. B}\ }\textbf {\bibinfo {volume} {90}},\ \bibinfo
  {pages} {174202} (\bibinfo {year} {2014})}\BibitemShut {NoStop}%
\bibitem [{\citenamefont {Luitz}\ \emph {et~al.}(2016)\citenamefont {Luitz},
  \citenamefont {Laflorencie},\ and\ \citenamefont {Alet}}]{luitz16}%
  \BibitemOpen
  \bibfield  {author} {\bibinfo {author} {\bibfnamefont {D.~J.}\ \bibnamefont
  {Luitz}}, \bibinfo {author} {\bibfnamefont {N.}~\bibnamefont {Laflorencie}},\
  and\ \bibinfo {author} {\bibfnamefont {F.}~\bibnamefont {Alet}},\ }\bibfield
  {title} {\bibinfo {title} {{Extended slow dynamical regime prefiguring the
  many-body localization transition}},\ }\href
  {https://doi.org/10.1103/PhysRevB.93.060201} {\bibfield  {journal} {\bibinfo
  {journal} {Phys. Rev. B}\ }\textbf {\bibinfo {volume} {93}},\ \bibinfo
  {pages} {060201} (\bibinfo {year} {2016})}\BibitemShut {NoStop}%
\bibitem [{\citenamefont {Mierzejewski}\ \emph {et~al.}(2016)\citenamefont
  {Mierzejewski}, \citenamefont {Herbrych},\ and\ \citenamefont
  {Prelov\v{s}ek}}]{mierzejewski16}%
  \BibitemOpen
  \bibfield  {author} {\bibinfo {author} {\bibfnamefont {M.}~\bibnamefont
  {Mierzejewski}}, \bibinfo {author} {\bibfnamefont {J.}~\bibnamefont
  {Herbrych}},\ and\ \bibinfo {author} {\bibfnamefont {P.}~\bibnamefont
  {Prelov\v{s}ek}},\ }\bibfield  {title} {\bibinfo {title} {{Universal dynamics
  of density correlations at the transition to many-body localized state}},\
  }\href {https://doi.org/10.1103/PhysRevB.94.224207} {\bibfield  {journal}
  {\bibinfo  {journal} {Phys. Rev. B}\ }\textbf {\bibinfo {volume} {94}},\
  \bibinfo {pages} {224207} (\bibinfo {year} {2016})}\BibitemShut {NoStop}%
\bibitem [{\citenamefont {\v{S}untajs}\ \emph
  {et~al.}(2020{\natexlab{b}})\citenamefont {\v{S}untajs}, \citenamefont
  {Bon\v{c}a}, \citenamefont {Prosen},\ and\ \citenamefont
  {Vidmar}}]{suntajs_bonca_20}%
  \BibitemOpen
  \bibfield  {author} {\bibinfo {author} {\bibfnamefont {J.}~\bibnamefont
  {\v{S}untajs}}, \bibinfo {author} {\bibfnamefont {J.}~\bibnamefont
  {Bon\v{c}a}}, \bibinfo {author} {\bibfnamefont {T.}~\bibnamefont {Prosen}},\
  and\ \bibinfo {author} {\bibfnamefont {L.}~\bibnamefont {Vidmar}},\
  }\bibfield  {title} {\bibinfo {title} {{Ergodicity breaking transition in
  finite disordered spin chains}},\ }\href
  {https://doi.org/10.1103/PhysRevB.102.064207} {\bibfield  {journal} {\bibinfo
   {journal} {Phys. Rev. B}\ }\textbf {\bibinfo {volume} {102}},\ \bibinfo
  {pages} {064207} (\bibinfo {year} {2020}{\natexlab{b}})}\BibitemShut
  {NoStop}%
\bibitem [{\citenamefont {Sels}\ and\ \citenamefont
  {Polkovnikov}(2021)}]{sels2020}%
  \BibitemOpen
  \bibfield  {author} {\bibinfo {author} {\bibfnamefont {D.}~\bibnamefont
  {Sels}}\ and\ \bibinfo {author} {\bibfnamefont {A.}~\bibnamefont
  {Polkovnikov}},\ }\bibfield  {title} {\bibinfo {title} {{Dynamical
  obstruction to localization in a disordered spin chain}},\ }\href
  {https://doi.org/10.1103/PhysRevE.104.054105} {\bibfield  {journal} {\bibinfo
   {journal} {Phys. Rev. E}\ }\textbf {\bibinfo {volume} {104}},\ \bibinfo
  {pages} {054105} (\bibinfo {year} {2021})}\BibitemShut {NoStop}%
\bibitem [{\citenamefont {Vidmar}\ \emph {et~al.}(2021)\citenamefont {Vidmar},
  \citenamefont {Krajewski}, \citenamefont {Bon\v{c}a},\ and\ \citenamefont
  {Mierzejewski}}]{vidmar21}%
  \BibitemOpen
  \bibfield  {author} {\bibinfo {author} {\bibfnamefont {L.}~\bibnamefont
  {Vidmar}}, \bibinfo {author} {\bibfnamefont {B.}~\bibnamefont {Krajewski}},
  \bibinfo {author} {\bibfnamefont {J.}~\bibnamefont {Bon\v{c}a}},\ and\
  \bibinfo {author} {\bibfnamefont {M.}~\bibnamefont {Mierzejewski}},\
  }\bibfield  {title} {\bibinfo {title} {{Phenomenology of Spectral Functions
  in Disordered Spin Chains at Infinite Temperature}},\ }\href
  {https://doi.org/10.1103/PhysRevLett.127.230603} {\bibfield  {journal}
  {\bibinfo  {journal} {Phys. Rev. Lett.}\ }\textbf {\bibinfo {volume} {127}},\
  \bibinfo {pages} {230603} (\bibinfo {year} {2021})}\BibitemShut {NoStop}%
\bibitem [{\citenamefont {Krajewski}\ \emph
  {et~al.}(2022{\natexlab{a}})\citenamefont {Krajewski}, \citenamefont
  {Vidmar}, \citenamefont {Bon\v{c}a},\ and\ \citenamefont
  {Mierzejewski}}]{krajewski22}%
  \BibitemOpen
  \bibfield  {author} {\bibinfo {author} {\bibfnamefont {B.}~\bibnamefont
  {Krajewski}}, \bibinfo {author} {\bibfnamefont {L.}~\bibnamefont {Vidmar}},
  \bibinfo {author} {\bibfnamefont {J.}~\bibnamefont {Bon\v{c}a}},\ and\
  \bibinfo {author} {\bibfnamefont {M.}~\bibnamefont {Mierzejewski}},\
  }\bibfield  {title} {\bibinfo {title} {{Restoring Ergodicity in a Strongly
  Disordered Interacting Chain}},\ }\href
  {https://doi.org/10.1103/PhysRevLett.129.260601} {\bibfield  {journal}
  {\bibinfo  {journal} {Phys. Rev. Lett.}\ }\textbf {\bibinfo {volume} {129}},\
  \bibinfo {pages} {260601} (\bibinfo {year} {2022}{\natexlab{a}})}\BibitemShut
  {NoStop}%
\bibitem [{\citenamefont {Bari\v{s}i\'{c}}\ \emph {et~al.}(2016)\citenamefont
  {Bari\v{s}i\'{c}}, \citenamefont {Kokalj}, \citenamefont {Balog},\ and\
  \citenamefont {Prelov\v{s}ek}}]{barisic16}%
  \BibitemOpen
  \bibfield  {author} {\bibinfo {author} {\bibfnamefont {O.~S.}\ \bibnamefont
  {Bari\v{s}i\'{c}}}, \bibinfo {author} {\bibfnamefont {J.}~\bibnamefont
  {Kokalj}}, \bibinfo {author} {\bibfnamefont {I.}~\bibnamefont {Balog}},\ and\
  \bibinfo {author} {\bibfnamefont {P.}~\bibnamefont {Prelov\v{s}ek}},\
  }\bibfield  {title} {\bibinfo {title} {{Dynamical conductivity and its
  fluctuations along the crossover to many-body localization}},\ }\href
  {https://doi.org/10.1103/PhysRevB.94.045126} {\bibfield  {journal} {\bibinfo
  {journal} {Phys. Rev. B}\ }\textbf {\bibinfo {volume} {94}},\ \bibinfo
  {pages} {045126} (\bibinfo {year} {2016})}\BibitemShut {NoStop}%
\bibitem [{\citenamefont {Prelov\v{s}ek}\ \emph {et~al.}(2021)\citenamefont
  {Prelov\v{s}ek}, \citenamefont {Mierzejewski}, \citenamefont {Krsnik},\ and\
  \citenamefont {Bari\v{s}\'{c}}}]{prelovsek21}%
  \BibitemOpen
  \bibfield  {author} {\bibinfo {author} {\bibfnamefont {P.}~\bibnamefont
  {Prelov\v{s}ek}}, \bibinfo {author} {\bibfnamefont {M.}~\bibnamefont
  {Mierzejewski}}, \bibinfo {author} {\bibfnamefont {J.}~\bibnamefont
  {Krsnik}},\ and\ \bibinfo {author} {\bibfnamefont {O.~S.}\ \bibnamefont
  {Bari\v{s}\'{c}}},\ }\bibfield  {title} {\bibinfo {title} {{Many-body
  localization as a percolation phenomenon}},\ }\href
  {https://doi.org/10.1103/PhysRevB.103.045139} {\bibfield  {journal} {\bibinfo
   {journal} {Phys. Rev. B}\ }\textbf {\bibinfo {volume} {103}},\ \bibinfo
  {pages} {045139} (\bibinfo {year} {2021})}\BibitemShut {NoStop}%
\bibitem [{\citenamefont {Herbrych}\ \emph {et~al.}(2022)\citenamefont
  {Herbrych}, \citenamefont {Mierzejewski},\ and\ \citenamefont
  {Prelov\v{s}ek}}]{herbrych22}%
  \BibitemOpen
  \bibfield  {author} {\bibinfo {author} {\bibfnamefont {J.}~\bibnamefont
  {Herbrych}}, \bibinfo {author} {\bibfnamefont {M.}~\bibnamefont
  {Mierzejewski}},\ and\ \bibinfo {author} {\bibfnamefont {P.}~\bibnamefont
  {Prelov\v{s}ek}},\ }\bibfield  {title} {\bibinfo {title} {{Relaxation at
  different length scales in models of many-body localization}},\ }\href
  {https://doi.org/10.1103/PhysRevB.105.L081105} {\bibfield  {journal}
  {\bibinfo  {journal} {Phys. Rev. B}\ }\textbf {\bibinfo {volume} {105}},\
  \bibinfo {pages} {L081105} (\bibinfo {year} {2022})}\BibitemShut {NoStop}%
\bibitem [{\citenamefont {Krajewski}\ \emph
  {et~al.}(2022{\natexlab{b}})\citenamefont {Krajewski}, \citenamefont
  {Mierzejewski},\ and\ \citenamefont {Bon\v{c}a}}]{krajewski22-2}%
  \BibitemOpen
  \bibfield  {author} {\bibinfo {author} {\bibfnamefont {B.}~\bibnamefont
  {Krajewski}}, \bibinfo {author} {\bibfnamefont {M.}~\bibnamefont
  {Mierzejewski}},\ and\ \bibinfo {author} {\bibfnamefont {J.}~\bibnamefont
  {Bon\v{c}a}},\ }\bibfield  {title} {\bibinfo {title} {{Modeling
  sample-to-sample fluctuations of the gap ratio in finite disordered spin
  chains}},\ }\href {https://doi.org/10.1103/PhysRevB.106.014201} {\bibfield
  {journal} {\bibinfo  {journal} {Phys. Rev. B}\ }\textbf {\bibinfo {volume}
  {106}},\ \bibinfo {pages} {014201} (\bibinfo {year}
  {2022}{\natexlab{b}})}\BibitemShut {NoStop}%
\bibitem [{\citenamefont {Prelov\v{s}ek}\ \emph {et~al.}(2023)\citenamefont
  {Prelov\v{s}ek}, \citenamefont {Herbrych},\ and\ \citenamefont
  {Mierzejewski}}]{prelovsek23}%
  \BibitemOpen
  \bibfield  {author} {\bibinfo {author} {\bibfnamefont {P.}~\bibnamefont
  {Prelov\v{s}ek}}, \bibinfo {author} {\bibfnamefont {J.}~\bibnamefont
  {Herbrych}},\ and\ \bibinfo {author} {\bibfnamefont {M.}~\bibnamefont
  {Mierzejewski}},\ }\bibfield  {title} {\bibinfo {title} {{Slow diffusion and
  Thouless localization criterion in modulated spin chains}},\ }\href
  {https://doi.org/10.1103/PhysRevB.108.035106} {\bibfield  {journal} {\bibinfo
   {journal} {Phys. Rev. B}\ }\textbf {\bibinfo {volume} {108}},\ \bibinfo
  {pages} {035106} (\bibinfo {year} {2023})}\BibitemShut {NoStop}%
\bibitem [{\citenamefont {Iyer}\ \emph {et~al.}(2013)\citenamefont {Iyer},
  \citenamefont {Oganesyan}, \citenamefont {Refael},\ and\ \citenamefont
  {Huse}}]{iyer13}%
  \BibitemOpen
  \bibfield  {author} {\bibinfo {author} {\bibfnamefont {S.}~\bibnamefont
  {Iyer}}, \bibinfo {author} {\bibfnamefont {V.}~\bibnamefont {Oganesyan}},
  \bibinfo {author} {\bibfnamefont {G.}~\bibnamefont {Refael}},\ and\ \bibinfo
  {author} {\bibfnamefont {D.~A.}\ \bibnamefont {Huse}},\ }\bibfield  {title}
  {\bibinfo {title} {{Many-body localization in a quasiperiodic system}},\
  }\href {https://doi.org/10.1103/PhysRevB.87.134202} {\bibfield  {journal}
  {\bibinfo  {journal} {Phys. Rev. B}\ }\textbf {\bibinfo {volume} {87}},\
  \bibinfo {pages} {134202} (\bibinfo {year} {2013})}\BibitemShut {NoStop}%
\bibitem [{\citenamefont {Bar~Lev}\ \emph {et~al.}(2017)\citenamefont
  {Bar~Lev}, \citenamefont {Kennes}, \citenamefont {Kl\"{o}ckner},
  \citenamefont {Reichman},\ and\ \citenamefont {Karrasch}}]{barlev17}%
  \BibitemOpen
  \bibfield  {author} {\bibinfo {author} {\bibfnamefont {Y.}~\bibnamefont
  {Bar~Lev}}, \bibinfo {author} {\bibfnamefont {D.~M.}\ \bibnamefont {Kennes}},
  \bibinfo {author} {\bibfnamefont {C.}~\bibnamefont {Kl\"{o}ckner}}, \bibinfo
  {author} {\bibfnamefont {D.~R.}\ \bibnamefont {Reichman}},\ and\ \bibinfo
  {author} {\bibfnamefont {C.}~\bibnamefont {Karrasch}},\ }\bibfield  {title}
  {\bibinfo {title} {{Transport in quasiperiodic interacting systems: From
  superdiffusion to subdiffusion}},\ }\href
  {https://doi.org/10.1209/0295-5075/119/37003} {\bibfield  {journal} {\bibinfo
   {journal} {EPL (Europhysics Letters)}\ }\textbf {\bibinfo {volume} {119}},\
  \bibinfo {pages} {37003} (\bibinfo {year} {2017})}\BibitemShut {NoStop}%
\bibitem [{\citenamefont {Khemani}\ \emph {et~al.}(2017)\citenamefont
  {Khemani}, \citenamefont {Sheng},\ and\ \citenamefont {Huse}}]{khemani17}%
  \BibitemOpen
  \bibfield  {author} {\bibinfo {author} {\bibfnamefont {V.}~\bibnamefont
  {Khemani}}, \bibinfo {author} {\bibfnamefont {D.~N.}\ \bibnamefont {Sheng}},\
  and\ \bibinfo {author} {\bibfnamefont {D.~A.}\ \bibnamefont {Huse}},\
  }\bibfield  {title} {\bibinfo {title} {{Two Universality Classes for the
  Many-Body Localization Transition}},\ }\href
  {https://doi.org/10.1103/PhysRevLett.119.075702} {\bibfield  {journal}
  {\bibinfo  {journal} {Phys. Rev. Lett.}\ }\textbf {\bibinfo {volume} {119}},\
  \bibinfo {pages} {075702} (\bibinfo {year} {2017})}\BibitemShut {NoStop}%
\bibitem [{\citenamefont {Setiawan}\ \emph {et~al.}(2017)\citenamefont
  {Setiawan}, \citenamefont {Deng},\ and\ \citenamefont {Pixley}}]{setiawan17}%
  \BibitemOpen
  \bibfield  {author} {\bibinfo {author} {\bibfnamefont {F.}~\bibnamefont
  {Setiawan}}, \bibinfo {author} {\bibfnamefont {D.~L.}\ \bibnamefont {Deng}},\
  and\ \bibinfo {author} {\bibfnamefont {J.~H.}\ \bibnamefont {Pixley}},\
  }\bibfield  {title} {\bibinfo {title} {{Transport properties across the
  many-body localization transition in quasiperiodic and random systems}},\
  }\href {https://doi.org/10.1103/PhysRevB.96.104205} {\bibfield  {journal}
  {\bibinfo  {journal} {Phys. Rev. B}\ }\textbf {\bibinfo {volume} {96}},\
  \bibinfo {pages} {104205} (\bibinfo {year} {2017})}\BibitemShut {NoStop}%
\bibitem [{\citenamefont {\v{Z}nidari\v{c}}\ and\ \citenamefont
  {Ljubotina}(2018)}]{znidaric18}%
  \BibitemOpen
  \bibfield  {author} {\bibinfo {author} {\bibfnamefont {M.}~\bibnamefont
  {\v{Z}nidari\v{c}}}\ and\ \bibinfo {author} {\bibfnamefont {M.}~\bibnamefont
  {Ljubotina}},\ }\bibfield  {title} {\bibinfo {title} {{Interaction
  instability of localization in quasiperiodic systems}},\ }\href
  {https://doi.org/10.1073/pnas.1800589115} {\bibfield  {journal} {\bibinfo
  {journal} {Proc. Natl. Acad. Sci. USA}\ }\textbf {\bibinfo {volume} {115}},\
  \bibinfo {pages} {4595} (\bibinfo {year} {2018})}\BibitemShut {NoStop}%
\bibitem [{\citenamefont {Zhang}\ and\ \citenamefont {Yao}(2018)}]{zhang18}%
  \BibitemOpen
  \bibfield  {author} {\bibinfo {author} {\bibfnamefont {S.~X.}\ \bibnamefont
  {Zhang}}\ and\ \bibinfo {author} {\bibfnamefont {H.}~\bibnamefont {Yao}},\
  }\bibfield  {title} {\bibinfo {title} {{Universal Properties of Many-Body
  Localization Transitions in Quasiperiodic Systems}},\ }\href
  {https://doi.org/10.1103/PhysRevLett.121.206601} {\bibfield  {journal}
  {\bibinfo  {journal} {Phys. Rev. Lett.}\ }\textbf {\bibinfo {volume} {121}},\
  \bibinfo {pages} {206601} (\bibinfo {year} {2018})}\BibitemShut {NoStop}%
\bibitem [{\citenamefont {Aramthottil}\ \emph {et~al.}(2021)\citenamefont
  {Aramthottil}, \citenamefont {Chanda}, \citenamefont {Sierant},\ and\
  \citenamefont {Zakrzewski}}]{aramthottil21}%
  \BibitemOpen
  \bibfield  {author} {\bibinfo {author} {\bibfnamefont {A.~S.}\ \bibnamefont
  {Aramthottil}}, \bibinfo {author} {\bibfnamefont {T.}~\bibnamefont {Chanda}},
  \bibinfo {author} {\bibfnamefont {P.}~\bibnamefont {Sierant}},\ and\ \bibinfo
  {author} {\bibfnamefont {J.}~\bibnamefont {Zakrzewski}},\ }\bibfield  {title}
  {\bibinfo {title} {{Finite-size scaling analysis of the many-body
  localization transition in quasiperiodic spin chains}},\ }\href
  {https://doi.org/10.1103/PhysRevB.104.214201} {\bibfield  {journal} {\bibinfo
   {journal} {Phys. Rev. B}\ }\textbf {\bibinfo {volume} {104}},\ \bibinfo
  {pages} {214201} (\bibinfo {year} {2021})}\BibitemShut {NoStop}%
\bibitem [{\citenamefont {Sierant}\ and\ \citenamefont
  {Zakrzewski}(2022)}]{sierant22}%
  \BibitemOpen
  \bibfield  {author} {\bibinfo {author} {\bibfnamefont {P.}~\bibnamefont
  {Sierant}}\ and\ \bibinfo {author} {\bibfnamefont {J.}~\bibnamefont
  {Zakrzewski}},\ }\bibfield  {title} {\bibinfo {title} {{Challenges to
  observation of many-body localization}},\ }\href
  {https://doi.org/10.1103/PhysRevB.105.224203} {\bibfield  {journal} {\bibinfo
   {journal} {Phys. Rev. B}\ }\textbf {\bibinfo {volume} {105}},\ \bibinfo
  {pages} {224203} (\bibinfo {year} {2022})}\BibitemShut {NoStop}%
\bibitem [{\citenamefont {Schreiber}\ \emph {et~al.}(2015)\citenamefont
  {Schreiber}, \citenamefont {Hodgman}, \citenamefont {Bordia}, \citenamefont
  {L{\"{u}}schen}, \citenamefont {Fischer}, \citenamefont {Vosk}, \citenamefont
  {Altman}, \citenamefont {Schneider},\ and\ \citenamefont
  {Bloch}}]{schreiber15}%
  \BibitemOpen
  \bibfield  {author} {\bibinfo {author} {\bibfnamefont {M.}~\bibnamefont
  {Schreiber}}, \bibinfo {author} {\bibfnamefont {S.~S.}\ \bibnamefont
  {Hodgman}}, \bibinfo {author} {\bibfnamefont {P.}~\bibnamefont {Bordia}},
  \bibinfo {author} {\bibfnamefont {H.~P.}\ \bibnamefont {L{\"{u}}schen}},
  \bibinfo {author} {\bibfnamefont {M.~H.}\ \bibnamefont {Fischer}}, \bibinfo
  {author} {\bibfnamefont {R.}~\bibnamefont {Vosk}}, \bibinfo {author}
  {\bibfnamefont {E.}~\bibnamefont {Altman}}, \bibinfo {author} {\bibfnamefont
  {U.}~\bibnamefont {Schneider}},\ and\ \bibinfo {author} {\bibfnamefont
  {I.}~\bibnamefont {Bloch}},\ }\bibfield  {title} {\bibinfo {title}
  {{Observation of many-body localization of interacting fermions in a
  quasi-random optical lattice}},\ }\href
  {https://doi.org/10.1126/science.aaa7432} {\bibfield  {journal} {\bibinfo
  {journal} {Science}\ }\textbf {\bibinfo {volume} {349}},\ \bibinfo {pages}
  {842} (\bibinfo {year} {2015})}\BibitemShut {NoStop}%
\bibitem [{\citenamefont {L{\"{u}}schen}\ \emph {et~al.}(2017)\citenamefont
  {L{\"{u}}schen}, \citenamefont {Bordia}, \citenamefont {Scherg},
  \citenamefont {Alet}, \citenamefont {Altman}, \citenamefont {Schneider},\
  and\ \citenamefont {Bloch}}]{luschen17}%
  \BibitemOpen
  \bibfield  {author} {\bibinfo {author} {\bibfnamefont {H.~P.}\ \bibnamefont
  {L{\"{u}}schen}}, \bibinfo {author} {\bibfnamefont {P.}~\bibnamefont
  {Bordia}}, \bibinfo {author} {\bibfnamefont {S.}~\bibnamefont {Scherg}},
  \bibinfo {author} {\bibfnamefont {F.}~\bibnamefont {Alet}}, \bibinfo {author}
  {\bibfnamefont {E.}~\bibnamefont {Altman}}, \bibinfo {author} {\bibfnamefont
  {U.}~\bibnamefont {Schneider}},\ and\ \bibinfo {author} {\bibfnamefont
  {I.}~\bibnamefont {Bloch}},\ }\bibfield  {title} {\bibinfo {title}
  {{Observation of Slow Dynamics near the Many-Body Localization Transition in
  One-Dimensional Quasiperiodic Systems}},\ }\href
  {https://doi.org/10.1103/PhysRevLett.119.260401} {\bibfield  {journal}
  {\bibinfo  {journal} {Phys. Rev. Lett.}\ }\textbf {\bibinfo {volume} {119}},\
  \bibinfo {pages} {260401} (\bibinfo {year} {2017})}\BibitemShut {NoStop}%
\bibitem [{\citenamefont {De~Roeck}\ \emph {et~al.}(2024)\citenamefont
  {De~Roeck}, \citenamefont {Giacomin}, \citenamefont {Huveneers},\ and\
  \citenamefont {Prosniak}}]{deroeck24}%
  \BibitemOpen
  \bibfield  {author} {\bibinfo {author} {\bibfnamefont {W.}~\bibnamefont
  {De~Roeck}}, \bibinfo {author} {\bibfnamefont {L.}~\bibnamefont {Giacomin}},
  \bibinfo {author} {\bibfnamefont {F.}~\bibnamefont {Huveneers}},\ and\
  \bibinfo {author} {\bibfnamefont {O.}~\bibnamefont {Prosniak}},\ }\bibfield
  {title} {\bibinfo {title} {{Absence of Normal Heat Conduction in Strongly
  Disordered Interacting Quantum Chains}},\ }\href@noop {} {\bibfield
  {journal} {\bibinfo  {journal} {arXiv}\ } (\bibinfo {year} {2024})},\ \Eprint
  {https://arxiv.org/abs/2408.04338} {arXiv:2408.04338} \BibitemShut {NoStop}%
\bibitem [{\citenamefont {Gopalakrishnan}\ \emph {et~al.}(2016)\citenamefont
  {Gopalakrishnan}, \citenamefont {Agarwal}, \citenamefont {Demler},
  \citenamefont {Huse},\ and\ \citenamefont {Knap}}]{gopal16}%
  \BibitemOpen
  \bibfield  {author} {\bibinfo {author} {\bibfnamefont {S.}~\bibnamefont
  {Gopalakrishnan}}, \bibinfo {author} {\bibfnamefont {K.}~\bibnamefont
  {Agarwal}}, \bibinfo {author} {\bibfnamefont {E.~A.}\ \bibnamefont {Demler}},
  \bibinfo {author} {\bibfnamefont {D.~A.}\ \bibnamefont {Huse}},\ and\
  \bibinfo {author} {\bibfnamefont {M.}~\bibnamefont {Knap}},\ }\bibfield
  {title} {\bibinfo {title} {{Griffiths effects and slow dynamics in nearly
  many-body localized systems}},\ }\href
  {https://doi.org/10.1103/PhysRevB.93.134206} {\bibfield  {journal} {\bibinfo
  {journal} {Phys. Rev. B}\ }\textbf {\bibinfo {volume} {93}},\ \bibinfo
  {pages} {134206} (\bibinfo {year} {2016})}\BibitemShut {NoStop}%
\bibitem [{\citenamefont {Jen\v{c}i\v{c}}\ and\ \citenamefont
  {Prelov{\v{s}}ek}(2015)}]{jencic15}%
  \BibitemOpen
  \bibfield  {author} {\bibinfo {author} {\bibfnamefont {B.}~\bibnamefont
  {Jen\v{c}i\v{c}}}\ and\ \bibinfo {author} {\bibfnamefont {P.}~\bibnamefont
  {Prelov{\v{s}}ek}},\ }\bibfield  {title} {\bibinfo {title} {{Spin and thermal
  conductivity in classical disordered spin chain}},\ }\href
  {https://doi.org/10.1103/PhysRevB.92.134305} {\bibfield  {journal} {\bibinfo
  {journal} {Phys. Rev. B}\ }\textbf {\bibinfo {volume} {92}},\ \bibinfo
  {pages} {134305} (\bibinfo {year} {2015})}\BibitemShut {NoStop}%
\bibitem [{\citenamefont {Mendoza-Arenas}\ \emph {et~al.}(2019)\citenamefont
  {Mendoza-Arenas}, \citenamefont {\v{Z}nidari\v{c}}, \citenamefont {Varma},
  \citenamefont {Goold}, \citenamefont {Clark},\ and\ \citenamefont
  {Scardicchio}}]{mendoza19}%
  \BibitemOpen
  \bibfield  {author} {\bibinfo {author} {\bibfnamefont {J.~J.}\ \bibnamefont
  {Mendoza-Arenas}}, \bibinfo {author} {\bibfnamefont {M.}~\bibnamefont
  {\v{Z}nidari\v{c}}}, \bibinfo {author} {\bibfnamefont {V.~K.}\ \bibnamefont
  {Varma}}, \bibinfo {author} {\bibfnamefont {J.}~\bibnamefont {Goold}},
  \bibinfo {author} {\bibfnamefont {S.~R.}\ \bibnamefont {Clark}},\ and\
  \bibinfo {author} {\bibfnamefont {A.}~\bibnamefont {Scardicchio}},\
  }\bibfield  {title} {\bibinfo {title} {{Asymmetry in energy versus spin
  transport in certain interacting disordered systems}},\ }\href
  {https://doi.org/10.1103/PhysRevB.99.094435} {\bibfield  {journal} {\bibinfo
  {journal} {Phys. Rev. B}\ }\textbf {\bibinfo {volume} {99}},\ \bibinfo
  {pages} {094435} (\bibinfo {year} {2019})}\BibitemShut {NoStop}%
\bibitem [{\citenamefont {Varma}\ \emph {et~al.}(2017)\citenamefont {Varma},
  \citenamefont {Lerose}, \citenamefont {Pietracaprina}, \citenamefont
  {Goold},\ and\ \citenamefont {Scardicchio}}]{varma17}%
  \BibitemOpen
  \bibfield  {author} {\bibinfo {author} {\bibfnamefont {V.~K.}\ \bibnamefont
  {Varma}}, \bibinfo {author} {\bibfnamefont {A.}~\bibnamefont {Lerose}},
  \bibinfo {author} {\bibfnamefont {F.}~\bibnamefont {Pietracaprina}}, \bibinfo
  {author} {\bibfnamefont {J.}~\bibnamefont {Goold}},\ and\ \bibinfo {author}
  {\bibfnamefont {A.}~\bibnamefont {Scardicchio}},\ }\bibfield  {title}
  {\bibinfo {title} {{Energy diffusion in the ergodic phase of a many body
  localizable spin chain}},\ }\href {https://doi.org/10.1088/1742-5468/aa668b}
  {\bibfield  {journal} {\bibinfo  {journal} {J. Stat. Mech. Theor. Exp.}\
  }\textbf {\bibinfo {volume} {2017}},\ \bibinfo {pages} {053101} (\bibinfo
  {year} {2017})}\BibitemShut {NoStop}%
\bibitem [{\citenamefont {Schulz}\ \emph {et~al.}(2018)\citenamefont {Schulz},
  \citenamefont {Taylor}, \citenamefont {Hooley},\ and\ \citenamefont
  {Scardicchio}}]{schulz18}%
  \BibitemOpen
  \bibfield  {author} {\bibinfo {author} {\bibfnamefont {M.}~\bibnamefont
  {Schulz}}, \bibinfo {author} {\bibfnamefont {S.~R.}\ \bibnamefont {Taylor}},
  \bibinfo {author} {\bibfnamefont {C.~A.}\ \bibnamefont {Hooley}},\ and\
  \bibinfo {author} {\bibfnamefont {A.}~\bibnamefont {Scardicchio}},\
  }\bibfield  {title} {\bibinfo {title} {{Energy transport in a disordered spin
  chain with broken U(1) symmetry: Diffusion, subdiffusion, and many-body
  localization}},\ }\href {https://doi.org/10.1103/PhysRevB.98.180201}
  {\bibfield  {journal} {\bibinfo  {journal} {Phys. Rev. B}\ }\textbf {\bibinfo
  {volume} {98}},\ \bibinfo {pages} {180201} (\bibinfo {year}
  {2018})}\BibitemShut {NoStop}%
\bibitem [{\citenamefont {Serbyn}\ \emph {et~al.}(2017)\citenamefont {Serbyn},
  \citenamefont {Papi\'{c}},\ and\ \citenamefont {Abanin}}]{serbyn17}%
  \BibitemOpen
  \bibfield  {author} {\bibinfo {author} {\bibfnamefont {M.}~\bibnamefont
  {Serbyn}}, \bibinfo {author} {\bibfnamefont {Z.}~\bibnamefont {Papi\'{c}}},\
  and\ \bibinfo {author} {\bibfnamefont {D.~A.}\ \bibnamefont {Abanin}},\
  }\bibfield  {title} {\bibinfo {title} {{Thouless energy and multifractality
  across the many-body localization transition}},\ }\href
  {https://doi.org/10.1103/PhysRevB.96.104201} {\bibfield  {journal} {\bibinfo
  {journal} {Phys. Rev. B}\ }\textbf {\bibinfo {volume} {96}},\ \bibinfo
  {pages} {104201} (\bibinfo {year} {2017})}\BibitemShut {NoStop}%
\bibitem [{\citenamefont {Edwards}\ and\ \citenamefont
  {Thouless}(1972)}]{edwards72}%
  \BibitemOpen
  \bibfield  {author} {\bibinfo {author} {\bibfnamefont {J.~T.}\ \bibnamefont
  {Edwards}}\ and\ \bibinfo {author} {\bibfnamefont {D.~J.}\ \bibnamefont
  {Thouless}},\ }\bibfield  {title} {\bibinfo {title} {{Numerical studies of
  localization in structurally disordered systems}},\ }\href
  {https://doi.org/10.1088/0022-3719/5/8/007} {\bibfield  {journal} {\bibinfo
  {journal} {J. Phys. C: Solid State Phys.}\ }\textbf {\bibinfo {volume} {5}},\
  \bibinfo {pages} {807} (\bibinfo {year} {1972})}\BibitemShut {NoStop}%
\bibitem [{\citenamefont {Sierant}\ \emph {et~al.}(2025)\citenamefont
  {Sierant}, \citenamefont {Lewenstein}, \citenamefont {Scardicchio},
  \citenamefont {Vidmar},\ and\ \citenamefont {Zakrzewski}}]{sierant24}%
  \BibitemOpen
  \bibfield  {author} {\bibinfo {author} {\bibfnamefont {P.}~\bibnamefont
  {Sierant}}, \bibinfo {author} {\bibfnamefont {M.}~\bibnamefont {Lewenstein}},
  \bibinfo {author} {\bibfnamefont {A.}~\bibnamefont {Scardicchio}}, \bibinfo
  {author} {\bibfnamefont {L.}~\bibnamefont {Vidmar}},\ and\ \bibinfo {author}
  {\bibfnamefont {J.}~\bibnamefont {Zakrzewski}},\ }\bibfield  {title}
  {\bibinfo {title} {{Many-Body Localization in the Age of Classical
  Computing}},\ }\href {https://doi.org/10.1088/1361-6633/ad9756} {\bibfield
  {journal} {\bibinfo  {journal} {Rep. Prog. Phys.}\ }\textbf {\bibinfo
  {volume} {88}},\ \bibinfo {pages} {026502} (\bibinfo {year}
  {2025})}\BibitemShut {NoStop}%
\bibitem [{\citenamefont {Wilkinson}(1990)}]{wilkinson90}%
  \BibitemOpen
  \bibfield  {author} {\bibinfo {author} {\bibfnamefont {M.}~\bibnamefont
  {Wilkinson}},\ }\bibfield  {title} {\bibinfo {title} {{Diffusion and
  dissipation in complex quantum systems}},\ }\href
  {https://doi.org/10.1103/PhysRevA.41.4645} {\bibfield  {journal} {\bibinfo
  {journal} {Phys. Rev. A}\ }\textbf {\bibinfo {volume} {41}},\ \bibinfo
  {pages} {4645} (\bibinfo {year} {1990})}\BibitemShut {NoStop}%
\bibitem [{\citenamefont {D'Alessio}\ \emph {et~al.}(2016)\citenamefont
  {D'Alessio}, \citenamefont {Kafri}, \citenamefont {Polkovnikov},\ and\
  \citenamefont {Rigol}}]{dalessio16}%
  \BibitemOpen
  \bibfield  {author} {\bibinfo {author} {\bibfnamefont {L.}~\bibnamefont
  {D'Alessio}}, \bibinfo {author} {\bibfnamefont {Y.}~\bibnamefont {Kafri}},
  \bibinfo {author} {\bibfnamefont {A.}~\bibnamefont {Polkovnikov}},\ and\
  \bibinfo {author} {\bibfnamefont {M.}~\bibnamefont {Rigol}},\ }\bibfield
  {title} {\bibinfo {title} {{From quantum chaos and eigenstate thermalization
  to statistical mechanics and thermodynamics}},\ }\href
  {https://doi.org/10.1080/00018732.2016.1198134} {\bibfield  {journal}
  {\bibinfo  {journal} {Adv. Phys.}\ }\textbf {\bibinfo {volume} {65}},\
  \bibinfo {pages} {239} (\bibinfo {year} {2016})}\BibitemShut {NoStop}%
\bibitem [{\citenamefont {Mori}(1965)}]{mori65}%
  \BibitemOpen
  \bibfield  {author} {\bibinfo {author} {\bibfnamefont {H.}~\bibnamefont
  {Mori}},\ }\bibfield  {title} {\bibinfo {title} {{Transport, collective
  motion, and Brownian motion}},\ }\href {https://doi.org/10.1143/PTP.33.423}
  {\bibfield  {journal} {\bibinfo  {journal} {Prog. Theor. Phys.}\ }\textbf
  {\bibinfo {volume} {33}},\ \bibinfo {pages} {423} (\bibinfo {year}
  {1965})}\BibitemShut {NoStop}%
\bibitem [{\citenamefont {Forster}(1975)}]{forster75}%
  \BibitemOpen
  \bibfield  {author} {\bibinfo {author} {\bibfnamefont {D.}~\bibnamefont
  {Forster}},\ }\href {https://doi.org/10.1201/9780429493683} {\emph {\bibinfo
  {title} {Hydrodynamic Fluctuations, Broken Symmetry and Correlation
  Functions}}}\ (\bibinfo  {publisher} {Taylor - Francis Group, CRC Press},\
  \bibinfo {year} {1975})\BibitemShut {NoStop}%
\bibitem [{\citenamefont {Bon\v{c}a}\ and\ \citenamefont
  {Jakli\v{c}}(1995)}]{bonca95}%
  \BibitemOpen
  \bibfield  {author} {\bibinfo {author} {\bibfnamefont {J.}~\bibnamefont
  {Bon\v{c}a}}\ and\ \bibinfo {author} {\bibfnamefont {J.}~\bibnamefont
  {Jakli\v{c}}},\ }\bibfield  {title} {\bibinfo {title} {{Spin diffusion of the
  t-J model}},\ }\href {https://doi.org/10.1103/PhysRevB.51.16083} {\bibfield
  {journal} {\bibinfo  {journal} {Phys. Rev. B}\ }\textbf {\bibinfo {volume}
  {51}},\ \bibinfo {pages} {16083} (\bibinfo {year} {1995})}\BibitemShut
  {NoStop}%
\bibitem [{\citenamefont {Karahalios}\ \emph {et~al.}(2009)\citenamefont
  {Karahalios}, \citenamefont {Metavitsiadis}, \citenamefont {Zotos},
  \citenamefont {Gorczyca},\ and\ \citenamefont
  {Prelov{\v{s}}ek}}]{karahalios09}%
  \BibitemOpen
  \bibfield  {author} {\bibinfo {author} {\bibfnamefont {A.}~\bibnamefont
  {Karahalios}}, \bibinfo {author} {\bibfnamefont {A.}~\bibnamefont
  {Metavitsiadis}}, \bibinfo {author} {\bibfnamefont {X.}~\bibnamefont
  {Zotos}}, \bibinfo {author} {\bibfnamefont {A.}~\bibnamefont {Gorczyca}},\
  and\ \bibinfo {author} {\bibfnamefont {P.}~\bibnamefont {Prelov{\v{s}}ek}},\
  }\bibfield  {title} {\bibinfo {title} {{Finite-temperature transport in
  disordered Heisenberg chains}},\ }\href
  {https://doi.org/10.1103/PhysRevB.79.024425} {\bibfield  {journal} {\bibinfo
  {journal} {Phys. Rev. B}\ }\textbf {\bibinfo {volume} {79}},\ \bibinfo
  {pages} {024425} (\bibinfo {year} {2009})}\BibitemShut {NoStop}%
\bibitem [{\citenamefont {Kohn}(1964)}]{kohn64}%
  \BibitemOpen
  \bibfield  {author} {\bibinfo {author} {\bibfnamefont {W.}~\bibnamefont
  {Kohn}},\ }\bibfield  {title} {\bibinfo {title} {{Theory of the Insulating
  State}},\ }\href {https://doi.org/10.1103/PhysRev.133.A171} {\bibfield
  {journal} {\bibinfo  {journal} {Phys. Rev.}\ }\textbf {\bibinfo {volume}
  {133}},\ \bibinfo {pages} {A171} (\bibinfo {year} {1964})}\BibitemShut
  {NoStop}%
\bibitem [{\citenamefont {Castella}\ and\ \citenamefont
  {Zotos}(1996)}]{castella96}%
  \BibitemOpen
  \bibfield  {author} {\bibinfo {author} {\bibfnamefont {H.}~\bibnamefont
  {Castella}}\ and\ \bibinfo {author} {\bibfnamefont {X.}~\bibnamefont
  {Zotos}},\ }\bibfield  {title} {\bibinfo {title} {{Finite-temperature
  mobility of a particle coupled to a fermionic environment}},\ }\href
  {https://doi.org/10.1103/PhysRevB.54.4375} {\bibfield  {journal} {\bibinfo
  {journal} {Phys. Rev. B}\ }\textbf {\bibinfo {volume} {54}},\ \bibinfo
  {pages} {4375} (\bibinfo {year} {1996})}\BibitemShut {NoStop}%
\bibitem [{\citenamefont {Long}\ \emph {et~al.}(2003)\citenamefont {Long},
  \citenamefont {Prelov\v{s}ek}, \citenamefont {El~Shawish}, \citenamefont
  {Karadamoglou},\ and\ \citenamefont {Zotos}}]{long03}%
  \BibitemOpen
  \bibfield  {author} {\bibinfo {author} {\bibfnamefont {M.}~\bibnamefont
  {Long}}, \bibinfo {author} {\bibfnamefont {P.}~\bibnamefont {Prelov\v{s}ek}},
  \bibinfo {author} {\bibfnamefont {S.}~\bibnamefont {El~Shawish}}, \bibinfo
  {author} {\bibfnamefont {J.}~\bibnamefont {Karadamoglou}},\ and\ \bibinfo
  {author} {\bibfnamefont {X.}~\bibnamefont {Zotos}},\ }\bibfield  {title}
  {\bibinfo {title} {{Finite-temperature dynamical correlations using the
  microcanonical ensemble and the Lanczos algorithm}},\ }\href
  {https://doi.org/10.1103/PhysRevB.68.235106} {\bibfield  {journal} {\bibinfo
  {journal} {Phys. Rev. B}\ }\textbf {\bibinfo {volume} {68}},\ \bibinfo
  {pages} {235106} (\bibinfo {year} {2003})}\BibitemShut {NoStop}%
\bibitem [{\citenamefont {Prelov\v{s}ek}\ and\ \citenamefont
  {Bon\v{c}a}(2013)}]{prelovsek11}%
  \BibitemOpen
  \bibfield  {author} {\bibinfo {author} {\bibfnamefont {P.}~\bibnamefont
  {Prelov\v{s}ek}}\ and\ \bibinfo {author} {\bibfnamefont {J.}~\bibnamefont
  {Bon\v{c}a}},\ }\bibfield  {title} {\bibinfo {title} {{Ground State and
  Finite Temperature Lanczos Methods}},\ }in\ \href
  {https://doi.org/10.1007/978-3-642-35106-8} {\emph {\bibinfo {booktitle}
  {Strongly Correlated Systems - Numerical Methods}}},\ \bibinfo {editor}
  {edited by\ \bibinfo {editor} {\bibfnamefont {A.}~\bibnamefont {Avella}}\
  and\ \bibinfo {editor} {\bibfnamefont {F.}~\bibnamefont {Mancini}}}\
  (\bibinfo  {publisher} {Springer, Berlin},\ \bibinfo {year}
  {2013})\BibitemShut {NoStop}%
\bibitem [{\citenamefont {Gopalakrishnan}\ \emph {et~al.}(2015)\citenamefont
  {Gopalakrishnan}, \citenamefont {M\"{u}ller}, \citenamefont {Khemani},
  \citenamefont {Knap}, \citenamefont {Demler},\ and\ \citenamefont
  {Huse}}]{gopal15}%
  \BibitemOpen
  \bibfield  {author} {\bibinfo {author} {\bibfnamefont {S.}~\bibnamefont
  {Gopalakrishnan}}, \bibinfo {author} {\bibfnamefont {M.}~\bibnamefont
  {M\"{u}ller}}, \bibinfo {author} {\bibfnamefont {V.}~\bibnamefont {Khemani}},
  \bibinfo {author} {\bibfnamefont {M.}~\bibnamefont {Knap}}, \bibinfo {author}
  {\bibfnamefont {E.}~\bibnamefont {Demler}},\ and\ \bibinfo {author}
  {\bibfnamefont {D.~A.}\ \bibnamefont {Huse}},\ }\bibfield  {title} {\bibinfo
  {title} {{Low-frequency conductivity in many-body localized systems}},\
  }\href {https://doi.org/10.1103/PhysRevB.92.104202} {\bibfield  {journal}
  {\bibinfo  {journal} {Phys. Rev. B}\ }\textbf {\bibinfo {volume} {92}},\
  \bibinfo {pages} {104202} (\bibinfo {year} {2015})}\BibitemShut {NoStop}%
\bibitem [{\citenamefont {Prelov\v{s}ek}\ and\ \citenamefont
  {Herbrych}(2017)}]{prelovsek17a}%
  \BibitemOpen
  \bibfield  {author} {\bibinfo {author} {\bibfnamefont {P.}~\bibnamefont
  {Prelov\v{s}ek}}\ and\ \bibinfo {author} {\bibfnamefont {J.}~\bibnamefont
  {Herbrych}},\ }\bibfield  {title} {\bibinfo {title} {{Self-consistent
  approach to many-body localization and subdiffusion}},\ }\href
  {https://doi.org/10.1103/PhysRevB.96.035130} {\bibfield  {journal} {\bibinfo
  {journal} {Phys. Rev. B}\ }\textbf {\bibinfo {volume} {96}},\ \bibinfo
  {pages} {035130} (\bibinfo {year} {2017})}\BibitemShut {NoStop}%
\bibitem [{\citenamefont {Potter}\ \emph {et~al.}(2015)\citenamefont {Potter},
  \citenamefont {Vasseur},\ and\ \citenamefont {Parameswaran}}]{potter15}%
  \BibitemOpen
  \bibfield  {author} {\bibinfo {author} {\bibfnamefont {A.~C.}\ \bibnamefont
  {Potter}}, \bibinfo {author} {\bibfnamefont {R.}~\bibnamefont {Vasseur}},\
  and\ \bibinfo {author} {\bibfnamefont {S.~A.}\ \bibnamefont {Parameswaran}},\
  }\bibfield  {title} {\bibinfo {title} {{Universal properties of many-body
  delocalization transitions}},\ }\href
  {https://doi.org/10.1103/PhysRevX.5.031033} {\bibfield  {journal} {\bibinfo
  {journal} {Phys. Rev. X}\ }\textbf {\bibinfo {volume} {5}},\ \bibinfo {pages}
  {031033} (\bibinfo {year} {2015})}\BibitemShut {NoStop}%
\bibitem [{ope()}]{opendata}%
  \BibitemOpen
  \href@noop {} {}\bibinfo {note} {All data discussed in this paper are
  available at
  \url{https://github.com/jacekherbrych/DataRepository}}\BibitemShut {NoStop}%
\end{thebibliography}%

\end{document}